\documentclass[aps,showkeys,showpacs,notitlepage,superscriptaddress]{revtex4-1}
\pdfoutput=1
\usepackage{amsmath}
\usepackage{amssymb}
\usepackage{amsfonts}
\usepackage{graphicx}
\usepackage{hyperref}
\usepackage{color}
\usepackage{multirow}
\usepackage{mathtools}
\usepackage{booktabs}
\usepackage{amsthm}
\usepackage{algorithm}
\usepackage{algpseudocode}
\usepackage{mathtools}
\usepackage{adjustbox}
\usepackage{xcolor}
\usepackage{color}
\usepackage{url}
\usepackage{tikz-cd}
\usepackage{ulem}

\def\dim{\operatorname{dim}}
\def\im{\operatorname{Image}}
\def\ker{\operatorname{Ker}}

\def\R{\mathbb R}

\def\N{\mathbb N}

\hypersetup{pdftitle=Forman-Ricci curvature and Persistent homology of
unweighted complex networks}

\begin{document}
\title{Forman-Ricci curvature and Persistent homology of unweighted complex
networks}

\author{Indrava Roy}
\email{Correspondence to: indrava@imsc.res.in}
\affiliation{The Institute of Mathematical Sciences (IMSc), Homi Bhabha National
Institute (HBNI), Chennai 600113 India}
\author{Sudharsan Vijayaraghavan}
\affiliation{Department of Applied Mathematics and Computational Sciences, PSG
College of Technology, Coimbatore 641004 India}
\author{Sarath Jyotsna Ramaia}
\affiliation{Department of Applied Mathematics and Computational Sciences, PSG
College of Technology, Coimbatore 641004 India}
\author{Areejit Samal}
\email{Correspondence to: asamal@imsc.res.in}
\affiliation{The Institute of Mathematical Sciences (IMSc), Homi Bhabha National
Institute (HBNI), Chennai 600113 India}
\affiliation{Max Planck Institute for Mathematics in the Sciences, Leipzig 04103
Germany}

\begin{abstract}
We present the application of topological data analysis (TDA) to study unweighted complex networks
via their persistent homology. By endowing appropriate weights that capture the inherent topological
characteristics of such a network, we convert an unweighted network into a weighted one. Standard
TDA tools are then used to compute their persistent homology. To this end, we use two main
quantifiers: a local measure based on Forman's discretized version of Ricci curvature, and a global
measure based on edge betweenness centrality. We have employed these methods to study various model
and real-world networks. Our results show that persistent homology can be used to distinguish between
model and real networks with different topological properties.
\end{abstract}

\maketitle

\section{Introduction}

Recent advances in topological data analysis (TDA) \cite{Zomorodian2005,Edelsbrunner2008,Carlsson2009}
have made it a powerful tool in data science. TDA has lead to important applications in different areas
of science. For example, in astrophysics, TDA has be used for analysis of the Cosmic Microwave
Background (CMB) radiation data \cite{Pranav2016}; in imaging, TDA has been used for feature detection
in 3D gray-scale images \cite{Gunther2011}; in biology, TDA has been used for detection of breast cancer
type with high survival rates \cite{Nicolau2011} and understanding cell fate from single-cell RNA
sequencing data \cite{Rizvi2017}. The main tool in TDA is that of \textit{persistent homology}
\cite{Zomorodian2005,Edelsbrunner2008,Carlsson2009}, which has the power to detect the topology of the
underlying data. The field of algebraic topology \cite{Munkres2018} provides the basic mathematical
tool required for TDA, namely that of homology. The conceptual roots of persistent homology, however,
are in \textit{differential} topology, in particular Morse theory \cite{Edelsbrunner2008}.

Network science \cite{Watts1998,Barabasi1999,Albert2002,Newman2010,Barabasi2016}, on the other hand,
investigates the topological and dynamical properties of various complex networks, that encode
interactions between various agents in the natural as well as artificial setting. The ability to
understand and predict the nature of these interactions is a key challenge. Historically, graph theory
\cite{Bollobas1998,Newman2010,Barabasi2016} has provided the main tools and techniques for studying
such networks, via their graph representation. Although graph theory has provided significant insights
into such problems, recent studies have shown that such techniques do not adequately capture higher-order
interactions and correlations arising in networks \cite{De2007,Horak2009,Petri2013,Petri2014,Bianconi2015,
Wu2015,Sizemore2016,Courtney2017,Ritchie2017,Courtney2018,Kartun-Giles2019,Iacopini2019,Kannan2019}.
These higher-order phenomena can be encoded in \textit{hypergraph} \cite{Klamt2009,Zlatic2009} and
\textit{simplicial complex} \cite{De2007,Horak2009,Lee2012,Petri2013,Petri2014,Sizemore2016,Iacopini2019}
representations of networks. The tools of TDA are applicable to any simplicial complex and can be used
to determine the important topological characteristics of networks. In this work, we employ TDA to study
the persistent homology of unweighted and undirected simple graphs arising from model and real-world
networks.

Previous research in this direction have investigated the persistent homology of weighted and undirected
networks \cite{Petri2013,Petri2014}. The filtration scheme required to compute persistent homology in
weighted networks was then provided by the edge weights \cite{Petri2013,Petri2014}. However, this technique
is not immediately applicable to unweighted graphs due to the absence of edge weights. At present, due to
insufficient information, the interaction networks underlying many real-world complex systems are available
only as unweighted and undirected graphs. Examples of such unweighted and undirected real networks include
the Yeast protein interaction network \cite{Jeong2001}, the US Power Grid network \cite{Leskovec2007} and
the Euro road network \cite{Subelj2011}. In order to reveal the higher-order topological features of such
real-world networks, it is important to develop methods to study persistent homology in unweighted and
undirected networks. A simple way to devise such a method would be to transform the given unweighted graph
into an edge-weighted graph by assigning certain weights to all edges, and then, using the induced filtration
to compute persistent homology. However, \textit{a priori} it is not evident which edge weighting scheme
would capture the topological characteristics of different types of unweighted networks.

Previously, Horak \textit{et al.} \cite{Horak2009} used a dimension-based weighting scheme for unweighted
networks where the weights are simply the dimension of the simplices. In particular, Horak \textit{et al.}
assign all edges with the weight $+1$ to study persistent homology in unweighted networks. However, we have
recently shown that the dimension-based filtration scheme of Horak \textit{et al.}, though computationally
fast, may not be able to conclusively distinguish between various model networks \cite{Kannan2019}. In
recent work \cite{Kannan2019}, we gave another weighting method based on a \textit{discrete Morse function} as
introduced by Robin Forman \cite{Forman1998,Forman2002}, which assigns weights to each simplex in the clique
complex corresponding to the unweighted graph according to a global acyclicity constraint. This method
\cite{Kannan2019} simplifies the topological structure of the underlying simplicial complex, that leads to a
computationally efficient way to compute homology and persistent homology. Moreover, we also showed that
the persistent homology computed using this method was able to distinguish various unweighted model networks
having different topological characteristics, the difference being quantified by the averaged bottleneck
distance between the corresponding persistence diagrams \cite{Kannan2019}. A natural question then is whether
other choices of weights can also be used to distinguish such unweighted networks via persistent homology.

In the present contribution, we shall use both local and global network quantifiers for obtaining edge weighting
schemes to compute persistent homology, namely that of \textit{discrete Ricci curvature} \cite{Forman2003,
Sreejith2016,Sreejith2017,Samal2018}, also introduced by R. Forman \cite{Forman2003}, which plays the role of
local curvature in a discrete setting, and \textit{edge betweenness centrality} \cite{Freeman1977,Girvan2002,
Newman2010} which is an edge-based measure analogous to the classical betweenness centrality for vertices of
a graph. We shall show that the simpler methods introduced here to study persistent homology based on Forman-Ricci
curvature or edge betweenness centrality are also able to distinguish unweighted model networks like our recent
method \cite{Kannan2019} based on discrete Morse functions. However, note that the advantages in topological
simplification and computational efficiency that result from using a discrete Morse function are lost with
the simpler method presented here. Nevertheless, if the sole goal is to compute persistent homology in unweighted
networks, the weighting schemes presented here are likely to be much simpler to use in practice. In this context,
we have also applied our methods to study the persistent homology of some real-world networks. Note that our recent
method based on discrete Morse functions \cite{Kannan2019} and the simpler methods presented here based on
Forman-Ricci curvature or edge betweenness centrality are much better at distinguishing between different types of
model networks in comparison to dimension-based method of Horak \textit{et al.} \cite{Horak2009}.

The remainder of the paper is organized as follows. In the Theory section, we present an overview of the concepts
needed to study persistent homology in unweighted networks based on Forman-Ricci curvature and edge betweenness
centrality. In the Datasets section, we describe the model and real networks analyzed here. In the Results section,
we describe our new methods to study persistent homology in unweighted networks, and its application to both model
and real-world networks. In the last section, we conclude with a brief summary and future outlook.


\begin{figure*}
\includegraphics[width=.81\columnwidth]{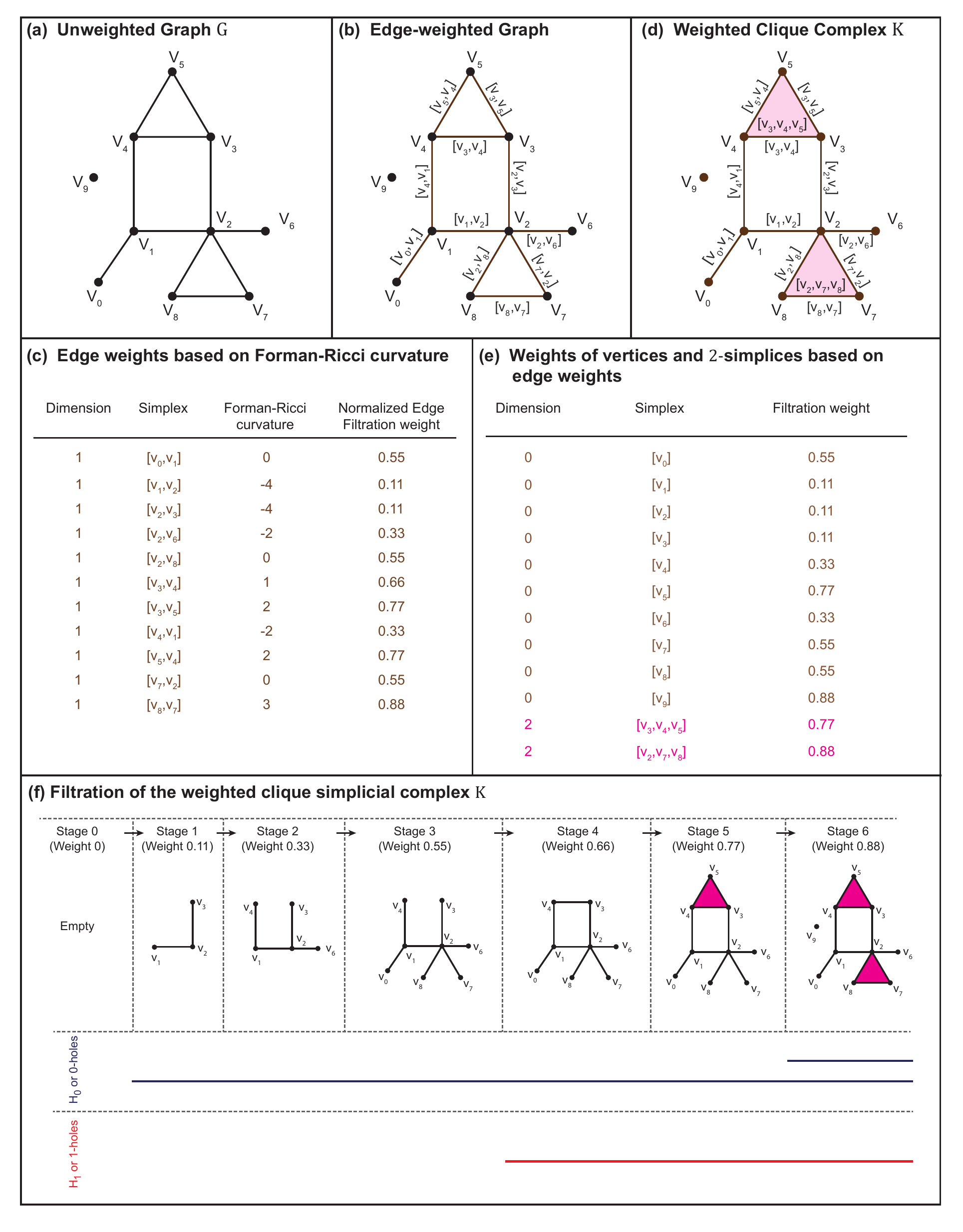}
\caption{Schematic figure illustrating our method to study persistent homology in an unweighted and
undirected network using Forman-Ricci curvature. (a) An example of an unweighted graph $G$. (b)
Transformation of the unweighted graph into an edge-weighted graph using Forman-Ricci curvature.
(c) Assignment of normalized filtration weights to edges in the weighted graph shown in (b) based
on Forman-Ricci curvature. (d) Weighted clique simplicial complex $K$ corresponding to the unweighted
graph $G$. (e) Assignment of normalized filtration weights to vertices ($0$-simplices) and $2$-simplices
in the weighted clique complex shown in (d) based on edge weights. (f) Filtration of the weighted
clique complex $K$ based on the ascending sequence of weights assigned to simplices. Barcodes depict
that there is a $0$-hole (or connected component) that persists across the 6 stages of the filtration
while another $0$-hole is born at the last stage on addition of the isolated vertex $v_9$. Moreover, a
$1$-hole is born at stage 4 on addition of the edge $[v_3,v_4]$.}
\label{schemfig}
\end{figure*}

\section{Theory}

\subsection{Clique complex of a graph}

Let $G(V,E)$ be a finite simple graph with $V$ being the set of vertices and $E$
being the set of edges. Each edge in the graph $G$ is an unordered pair of distinct
vertices. We remark that a simple graph does not contain self-loops or multi-edges
\cite{Bollobas1998}. An induced subgraph $K$ of $G$ that is complete is called a
\textit{clique}. We can view $G$ as a finite clique simplicial complex $K$ where a
$p$-dimensional simplex (or $p$-simplex) is determined by a set of $p+1$ vertices
that form a clique \cite{Zomorodian2005,Edelsbrunner2008}. Specifically, a $p$-simplex
is a polytope which is the convex hull of its $p+1$ vertices. Note that a \textit{simplex}
can be thought of as a generalization of points, lines, triangles, tetrahedron, and so on
in higher dimensions. In the clique complex $K$, $0$-simplices correspond to vertices
in $G$, $1$-simplices to edges in $G$, $2$-simplices to triangles in $G$, and so on.
Given a $p$-simplex $\alpha$ in $K$, a \textit{face} $\gamma$ of $\alpha$ is determined
by a subset of the vertex set of $\alpha$ of cardinality less than or equal to $p+1$.
Dually, a \textit{co-face} $\beta$ of $\alpha$ is a simplex that contains $\alpha$ as
a face. The dimension of a clique simplicial complex $K$ is the maximum dimension of
its constituent simplices. An orientation of a $p$-simplex is given by an ordering of its
constituent vertices \cite{Munkres2018}. Moreover, two orientations of a simplex are
equivalent if they differ by an even permutation of its vertices.


\subsection{Persistent homology of a simplicial complex}

A simplicial complex is a collection $K$ of simplices which satisfies following
two properties \cite{Munkres2018}. Firstly, any face $\gamma$ of a simplex $\alpha$
in $K$ is also included in $K$. Secondly, if two simplices $\alpha$ and $\beta$
in $K$ have a non-empty intersection $\gamma$, then $\gamma$ is a common face of
$\alpha$ and $\beta$. A subcomplex $K'$ of a simplicial complex $K$ is a collection
of simplices in $K$ such that $K'$ is also a simplicial complex. A \textit{filtration}
on a simplicial complex $K$ is given by a nested sequence of subcomplexes $K_i$,
$i=0,1,\ldots,n$, such that:
\begin{equation*}
\emptyset=K_0\subseteq K_1\subseteq \ldots \subseteq K_n=K.
\end{equation*}

For a simplicial complex $K$ with a given filtration, one can define its persistent
homology groups as follows. First we fix a base field $\mathbb{F}$ \cite{Munkres2018}.
The set of all oriented $p$-simplices in $K$ generate a free group $C_p$ over
$\mathbb{F}$, called $p^{\text{th}}$-chain group \cite{Munkres2018}. An element in
$C_p$ is called a $p$-chain, and is given by a finite formal sum:
\begin{equation*}
C_p =\sum_{i=1}^{N} c_i\alpha_i
\end{equation*}
where the coefficients $c_i$ are in $\mathbb{F}$, and $\alpha_i$ are oriented
$p$-simplices in $K$ \cite{Munkres2018}. Component-wise addition endows $C_p$ with the
structure of a group, whose identity element is given by the unique $p$-chain with all
coefficients $c_i$ equal to zero. If a $p$-simplex $\alpha$ is given an opposite
orientation, then it is represented as $-\alpha$ in $C_p$, and gives the inverse of
$\alpha$ in $C_p$. To define the persistent homology groups, we use the so-called boundary
operator $\partial_p$, which is a map $\partial_p: C_p\rightarrow C_{p-1}$.

For an oriented $p$-simplex $\alpha = [x_0,x_1,\ldots,x_p]$ (i.e., the ordered vertex set
$\{x_0, x_1,\ldots,x_p\}$ of $\alpha$), we define the boundary operator $\partial_p$ as:
\begin{equation*}
\partial_p(\alpha)=\sum_{i=0}^{p}(-1)^i[x_0,\ldots,\hat{x}_i, \ldots,x_p]
\end{equation*}
where $[x_0,\ldots,\hat{x}_i,\ldots,x_p]$ denotes the $(p-1)$-face of $\alpha$ obtained by
removing the vertex $x_i$ \cite{Munkres2018}. Since the right hand side of the above
equation is a linear combination of $(p-1)$-simplices, it belongs to $C_{p-1}$. One can
then extend the definition of $\partial_p$ to all elements of $C_p$ by linearity. The
boundary operators satisfy the fundamental property:
\begin{equation*}
\partial_p\circ \partial_{p+1}=0.
\end{equation*}

The kernel of the boundary operator is called the group of $p$-cycles and denoted by $Z_p$
\cite{Munkres2018}. It is given by the set of elements in $C_p$ that is mapped to $0$ in
$C_{p-1}$ by the boundary operator $\partial_p$:
\begin{equation*}
Z_p =\ker(\partial_p)=\{c\in C_p|\partial_p(c)=0\}.
\end{equation*}

A $p$-boundary is a $p$-cycle which lies in the image of the boundary operator $\partial_{p+1}$.
The set of $p$-boundaries is denoted by $B_p$ and is a subgroup of $Z_p$ \cite{Munkres2018}.
\begin{equation*}
B_p =\im(\partial_{p+1})=\{c\in  C_p| \exists b \in C_{p+1},
\partial_{p+1}(b)=c\}.
\end{equation*}

Thus, the $p$-homology group is defined as \cite{Munkres2018}:
\begin{equation*}
H_p(K)=\frac{Z_p(K)}{B_p(K)}.
\end{equation*}
Note that $H_p$ is a vector space over the field $\mathbb{F}$. The $p$-Betti number $\beta_p$ is given by the
dimension of the homology group $H_p$. Informally, $\beta_p$ represents the number of holes in the $p$-homology group.

Now, every subcomplex $K_i$ in the filtration of the simplicial complex $K$ has an index $i$
associated with it. Also, for each $K_i$ there exists its corresponding $p$-chain, $p$-boundary
operators, and thus, $p$-boundaries and $p$-cycles. We shall denote the $p$-cycles of $K_i$ as
$Z^i_p$ and the $p$-boundaries of $K_i$ as $B^i_p$. The $j$-persistent $p$-homology of $K_i$ is
defined as \cite{Zomorodian2005,Edelsbrunner2008}:
\begin{equation}
H^{i,j}_p = \frac{Z^i_p}{(B^{i+j}_p \cap Z^i_p)}
\end{equation}
and the corresponding $j$-persistent $p$-Betti number as:
\begin{equation}
\beta^{i,j}_p = \dim(H^{i,j}_p).
\end{equation}

A $p$-homology class $\alpha$ is \textit{born} at $K_i$ if it is not in the image of the map
induced on $p$-homology by the inclusion $K_{i-1} \subseteq K_i$ . Furthermore, if $\alpha$ is
born at $K_i$, we say that it \textit{dies} entering $K_{i+j}$, if it becomes the boundary of a
$(p+1)$-chain in $K_{i+j}$. The persistent homology group $H^{i,j}_p$ thus encodes information of
$p$-homology classes that are born at the filtration index $i$ and survive until the index $i+j$.
Each $p$-hole across the filtration can be characterized by its birth and death. By studying
persistent homology, the persistence of such holes can be quantified, thus revealing the
importance of the corresponding topological features across the filtration.

\subsubsection{Filtration of a weighted simplicial complex}

Let $K$ be a simplicial complex, endowed with a set of numerical numbers called \textit{weights}
associated with its simplices, i.e. to each of its constituent simplices $\alpha$ is assigned a
number $w(\alpha)$. To study the persistent homology of such simplicial complexes, we can consider
the following filtration on $K$ \cite{Zomorodian2005,Edelsbrunner2008}. Given a real number $r$,
we define the subcomplex $K(r)$ as:
\begin{equation}
K(r)= \bigcup_{\{\alpha: w(\alpha)\leq r\}} \bigcup_{\beta\leq \alpha} \beta
\end{equation}
In simple terms, $K(r)$ is the smallest simplicial subcomplex of $K$ containing simplices which have
weight less than or equal to $r$. Note that all faces $\beta$ of an simplex $\alpha$ of weight less
than $r$ are admitted to this subcomplex, irrespective of the weight of $\beta$. In the particular
case of a finite simplicial complex $K$ with simplices $\alpha_i, i=1,2,\ldots,n$, we can arrange the
corresponding weights $w(\alpha_i)$ in ascending order, say $\lambda_1 \leq \lambda_2 \leq \cdots \leq
\lambda_n$. Then the associated filtration of $K$ is given by:
\begin{equation}
\label{filtration}
\emptyset\subseteq K(\lambda_1)\subseteq K(\lambda_2)\subseteq \ldots \subseteq K(\lambda_n)=K
\end{equation}

Throughout this work, we shall consider the particular case of a finite weighted simplicial complex
arising from a finite \textit{edge-weighted} simple graph, i.e. a graph with weights assigned to all
its edges. The clique complex of such an edge-weighted graph already has weights on its $1$-simplices
corresponding to its edges, and we can extend this weighting scheme to any $0$-simplex $\beta$ or
$2$-simplex $\sigma$ by defining their weights $w(\beta)$ and $w(\sigma)$ by the following
\textit{min/max} formulae:
\begin{eqnarray}
w(\beta)= \min\{ w(\alpha) : \alpha \text{ is a 1-dimensional co-face of } \beta\}\ \text{and} \nonumber \\
w(\sigma)= \max\{ w(\alpha) : \alpha \text{ is a 1-dimensional face of } \sigma\}.
\label{simplexweight}
\end{eqnarray}
Weights of higher-dimensional simplices can be defined in a similar way. Note that with these weights,
we enforce the following conditions. Any vertex of an edge $e$ that is included in a subcomplex containing
$e$ has weight less than or equal to $e$. Moreover, if a collection of edges forms a higher-dimensional
simplex $\gamma$, then $\gamma$ is included in a subcomplex that includes the edge with the maximum weight.
With respect to the filtration induced by this weighting scheme arranged in increasing order
$\lambda_1 \leq \lambda_2 \leq \cdots \leq \lambda_n$, one can now compute the persistent homology
groups of $K$ as defined above. The \textit{persistence} of a class in $p$-homology that is born at
the $i^{\text{th}}$-stage of the filtration and dies at the $j^{\text{th}}$-stage is then defined to be
$\lambda_j- \lambda_i$, where $\lambda_i$ is the weight associated to the $i^{\text{th}}$-subcomplex
$K(\lambda_i)$ as above.

\subsubsection{Barcode diagrams}

The $p^{\text{th}}$-barcode diagram for a given filtration of a finite simplicial complex $K$ gives a
graphical summary of the birth and death of $p$-holes across the filtration \cite{Ghrist2008}. In this
work, the x-axis of the $p^{\text{th}}$-barcode diagram corresponds to the filtration weights of
$p$-simplices in $K$; the filtration weights have been normalized to lie in the range 0 to 1. A horizontal
line in the $p^{\text{th}}$-barcode diagram of $K$ is referred to as a barcode. A barcode that begins at a
x-axis value of $w_i$ and ends at a x-axis value of $w_j$ represents a $p$-hole in $K$ whose birth and death
weights are $w_i$ and $w_j$, respectively. The number of barcodes between $w_i$ and $w_j$ in the diagram
is precisely the $p$-Betti number $\beta^{i,j-i}$, i.e., the dimension of the persistent homology group
$H^{i,j-i}_p$.

\subsubsection{Persistence diagrams and bottleneck distance between them}

Given two multisets $X$ and $Y$ in $\mathbb{R}^2$, the $\infty$-Wasserstein distance or bottleneck distance
between them is defined as:
\begin{equation*}
\label{botdist}
W_\infty(X,Y) =  \inf_{\eta:X \rightarrow Y} \text{sup}_{x \in X} || x - \eta(x) ||_{\infty}.
\end{equation*}
In the above equation, the supremum is taken over all bijections $\eta:X\rightarrow Y$ (with the convention
that a point with multiplicity $k \in \N$ is considered as $k$ individual points) and for $(a,b) \in
\mathbb{R}^2$, $||(a,b)||_\infty:= \max\{|a|, |b|\}$ \cite{DiFabio2015}.

Given a filtration of a weighted simplicial complex $K$ with weights $w_i, i= 1, 2, \cdots, n$, the
$p^{\text{th}}$-persistence diagram $D^pK$, is defined as follows. Consider the multiset of points
$W^pK:=\{(w_i, w_j): w_i<w_j, i, j = 1, 2, \cdots, n\}$ with each point $(w_i, w_j)$ endowed with the
multiplicity $\mu_p(w_i, w_j)$ given by \cite{DiFabio2015}:
\begin{equation*}
\mu_p(w_i, w_j) := \lim_{\epsilon \rightarrow 0^+} (\beta_{w_i+\epsilon}^{w_j-\epsilon}
- \beta_{w_i+\epsilon}^{w_j+\epsilon} + \beta_{w_i-\epsilon}^{w_j+\epsilon}
- \beta_{w_i-\epsilon}^{w_j-\epsilon})
\end{equation*}
where $\beta_x^y$ is the dimension of the image of the induced map in $p$-homology from $K(x)$ to $K(y)$ for
$x, y \in \R$ with $x<y$. Denote by $\Delta$ the diagonal in $\R^2$ considered as a multiset with infinite
multiplicity given to each of its points. The persistence diagram $D^pK$ is the subset of $W^p_K \cup \Delta$
consisting of points $(u, v)$ with $\mu_p(u, v)>0$. In this work, we consider the total persistence diagram
\cite{Cohen-Steiner2007} given by the union of all $D^pK$ for $0 \leq p \leq \dim{K}$. Thereafter, we consider
the bottleneck distance \cite{Cohen-Steiner2007} between total persistence diagrams considered as multisets
in $\mathbb{R}^2$.


\subsection{Forman-Ricci curvature}

In previous work \cite{Sreejith2016,Sreejith2017,Samal2018}, some of us have
ported a discretization of the classical notion of Ricci curvature due to Robin
Forman \cite{Forman2003} to graphs or networks. Briefly, in Riemannian geometry,
curvature measures the amount of deviation of a smooth Riemannian manifold from
being Euclidean. The Ricci curvature tensor quantifies the dispersion of geodesic
lines in the neighbourhood of a given tangential direction as well as volume
growth of metric balls. Forman \cite{Forman2003} has proposed a discretization
of the classical Ricci curvature based on the \textit{Bochner-Weitzenb\"{o}ck
formula} which measures the difference between the \textit{Laplace-Beltrami
operator} and the \textit{connection Laplacian} \cite{Jost2017}. Forman's
discretized version of the Ricci curvature is applicable to a large class of
topological objects, namely, \textit{weighted $CW$-complexes} which includes
graphs and simplicial complexes \cite{Forman2003,Sreejith2016}.

Starting from a graph or network, one may construct a two-dimensional polyhedral
complex by inserting a solid triangle into any connected triple of vertices or
cycle of length 3, a solid quadrangle into a cycle of length 4, a solid pentagon
into a cycle of length 5, and so on. The mathematical definition of Forman-Ricci
curvature \cite{Forman2003} for general weighted $CW$-complexes is also applicable
to such a two-dimensional polyhedral complex constructed from a graph, and is
given by:
\begin{equation*}
{\rm F} (e) = w_e \left[ \sum_{e < f} \frac{w_e}{w_f}+\sum_{v < e}
\frac{w_v}{w_e} \right.
- \left. \sum_{\hat{e} \parallel e} \left| \sum_{\hat{e},e < f} \frac{\sqrt{w_e
\cdot w_{\hat{e}}}}{w_f}
- \sum_{v 	 < e, v < \hat{e}} \frac{w_v}{\sqrt{w_e \cdot w_{\hat{e}}}}
\right| \right] \; ;
\end{equation*}
where $w_e$ denotes the weight of edge $e$, $w_v$ denotes the weight of vertex
$v$, $w_f$ denotes the weight of face $f$, $\sigma < \tau$ means that $\sigma$
is a face of $\tau$, and $||$ signifies \textit{parallelism}, i.e. the two cells
have a common \textit{parent} (higher dimensional co-face) or a common
\textit{child} (lower dimensional face), but not both a common parent and common
child \cite{Samal2018}. For the particular case of restricted two-dimensional
complexes containing only triangular faces $t$ while ignoring faces consisting
of more than 3 vertices, the above equation simplifies to \cite{Samal2018}:
\begin{equation}
\label{AugmentedFormanRicciEdge}
{\rm F} (e) = w_e \left[ \sum_{e < t} \frac{w_e}{w_t} + (\frac{w_{v_1}}{w_e} +
\frac{w_{v_2}}{w_e}) \right.
- \left. \sum_{e_{v_1},e_{v_2}\ \sim\ e,\ e_{v_1},e_{v_2}\ \nless\ t} \left(
\frac{w_{v_1}}{\sqrt{w_e \cdot w_{e_{v_1}}}}
+ \frac{w_{v_2}}{\sqrt{w_e \cdot w_{e_{v_2}}}} \right) \right] \; ;
\end{equation}
where $w_e$ is the weight of the edge $e$ under consideration, $w_{v_1}$ and
$w_{v_2}$ denote the weights associated with the vertices $v_1$ and $v_2$,
respectively, which anchor the edge $e$ under consideration. In the above
equation, $e_{v_1} \sim e$ and $e_{v_2} \sim e$ denote the set of edges incident
on vertices $v_1$ and $v_2$, respectively, after \textit{excluding} the edge
$e$ under consideration which connects the two vertices $v_1$ and $v_2$.
While computing the Forman-Ricci curvature of an edge in an unweighted graph
$G$, we substitute in the above equation $w_t = w_e = w_v = 1, \; \forall\ t
\in T(G), e \in E(G), v \in V(G)$, where $T(G)$, $E(G)$ and $V(G)$ represent
the set of triangular faces, edges and vertices, respectively. Note that the
above definition (Eq. \ref{AugmentedFormanRicciEdge}) of the Forman-Ricci
curvature of an edge or $1$-simplex in the restricted two-dimensional complex
constructed from a graph was referred to as \textit{augmented} Forman-Ricci
curvature in earlier contributions \cite{Samal2018,Saucan2019}. For brevity,
we here refer to the quantity defined in Eq. \ref{AugmentedFormanRicciEdge}
as Forman-Ricci curvature of an edge. From a geometric point of view, the
Forman-Ricci curvature quantifies the information spread at the ends of edges
in a network. Higher information spread at the ends of an edge implies more
negative value for its Forman-Ricci curvature. In this work, we employ
Forman-Ricci curvature of an edge (given by Eq. \ref{AugmentedFormanRicciEdge})
to transform an unweighted graph into a weighted graph which captures the
local curvature properties (Figure \ref{schemfig}).


\subsection{Edge Betweenness Centrality}

Edge betweenness centrality \cite{Freeman1977,Girvan2002,Newman2010} quantifies
the importance of edges for global information flow in networks. For any edge $e$,
this measure is computed based on the number of shortest paths between different
pairs of vertices in the network that contain the considered edge $e$. Formally,
in a graph $G(V,E)$, the edge betweenness centrality of an edge $e \in E$ is given
by:
\begin{equation}
\label{EBC}
{\rm EBC} (e) = \sum_{v_i} \sum_{v_j, v_j \neq v_i} \frac{\sigma_{v_i
v_j}(e)}{\sigma_{v_i v_j}}
\end{equation}
where $\sigma_{v_i v_j}$ gives the number of shortest paths between vertices
$v_i$ and $v_j$ in the network and $\sigma_{v_i v_j}(e)$ gives the number of
shortest paths between vertices $v_i$ and $v_j$ in the network that contain the
considered edge $e$. Note that an edge with a high edge betweenness centrality
is critical for maintaining information flow in the network.


\section{Datasets}

The proposed method for studying persistent homology in unweighted networks has
been investigated in different network models, namely, Erd\"{o}s-Renyi (ER) model
\cite{Erdos1961}, Watts-Strogatz (WS) model \cite{Watts1998}, Barab\'{a}si-Albert
(BA) model \cite{Barabasi1999}, and the Hyperbolic Graph Generator (HGG)
\cite{Krioukov2010}. We give brief descriptions of each below.
\begin{itemize}
\setlength\itemsep{0em}
\item \textit{ER model}: The ER model has two parameter $n$ and $p$, where $n$
is the number of vertices and $p$ is the probability for the existence of an edge
between distinct pairs of vertices. ER graph is obtained by starting with a set
of vertices and connecting a distinct pair of vertices by an edge with probability
$p$. The presence of an edge between any two pairs of vertices is independent of the
other edges.
\item \textit{WS model}: The WS model can be characterized by three parameters:
$n$, the number of vertices; $k$, the number of neighbours the vertex has before
rewiring; and $p$, the rewiring probability. The construction of the WS graph begins
with a graph with $n$ vertices where each vertex has $k$ nearest neighbours. Thereafter,
the endpoints of each edge is chosen randomly based on the rewiring probability and
it is rewired to another randomly chosen vertex with uniform probability.
\item \textit{BA model}: The BA model generates a scale-free graph with $n$ vertices
by satisfying the so-called preferential attachment condition. Under preferential
attachment scheme, at each iteration step, the graph expansion takes place by addition
of a new vertex with $m$ edges to existing vertices in such a way that existing vertices
with higher degree have more probability to gain additional edges to the new vertex
than the vertices with lower degree. In the BA model, the probability of connecting
the new vertex to an existing vertex is directly proportional to the degree of that
vertex at that time. The BA model generates graphs with power-law degree distribution
\cite{Barabasi1999}.
\item \textit{Hyperbolic random graphs}: The input parameters of HGG are the number of
vertices $n$, the target average degree $k$, the target exponent $\gamma$ of the power-law
degree distribution and temperature $T$. For the construction of a hyperbolic random graph,
HGG scatters $n$ vertices on a hyperbolic space and the existence of an edge between the
vertices is based on a probability value, which is given by a function of the hyperbolic
distance between the vertices. The vertex degree distribution in the hyperbolic random
graphs produced by HGG follow a power-law. The HGG generates a hyperbolic random graph for
$\gamma=[0,\infty)$.
\item \textit{Spherical random graphs}: Similar to the hyperbolic random graph,
a spherical random graph can be constructed using HGG by scattering $n$ vertices on a
sphere of radius $R$, and the probability for existence of an edge between two vertices
is a function of the spherical distance between the vertices. The HGG model produces
a spherical random graph for $\gamma=\infty$.
\end{itemize}

In addition to model networks, the proposed method has also been studied in the following
seven real-world networks.
\begin{itemize}
\setlength\itemsep{0em}
\item \textit{Yeast protein interaction} network \cite{Jeong2001} with 1870 vertices
representing proteins and 2277 edges signifying protein-protein interactions.
\item \textit{Human protein interaction} network \cite{Rual2005} with 3133 vertices
representing proteins and 6726 edges signifying protein-protein interactions.
\item \textit{US Power Grid} network \cite{Leskovec2007} with 4941 vertices representing
generators, transformers and substations in the Western states of USA and the 6594 edges
signifying power links between them.
\item \textit{Euro road} network \cite{Subelj2011} with 1174 vertices corresponding to
cities in Europe and the 1417 edges signifying roads linking the cities.
\item \textit{Email} network \cite{Guimera2003} with 1133 vertices representing users in
the University of Rovira i Virgili and 5451 edges signifying the existence of at least one
Email communication between pairs of users.
\item \textit{Route views} network \cite{Leskovec2007} with 6474 vertices corresponding
to autonomous systems and 13895 edges signifying communication between the autonomous systems
or vertices.
\item \textit{Hamsterster friendship} network \cite{Kunegis2013} with 1858 vertices
representing the users and 12534 edges signifying friendships between the users.
\end{itemize}
We remark that self-loops have been omitted while constructing the clique complexes from
the undirected graphs corresponding to real networks. Note that the dataset of model and
real-world networks analyzed here using the proposed methods based on Forman-Ricci curvature
or edge betweenness centrality were also studied using our previous method \cite{Kannan2019}
based on discrete Morse theory.


\begin{figure*}
\includegraphics[width=.7\columnwidth]{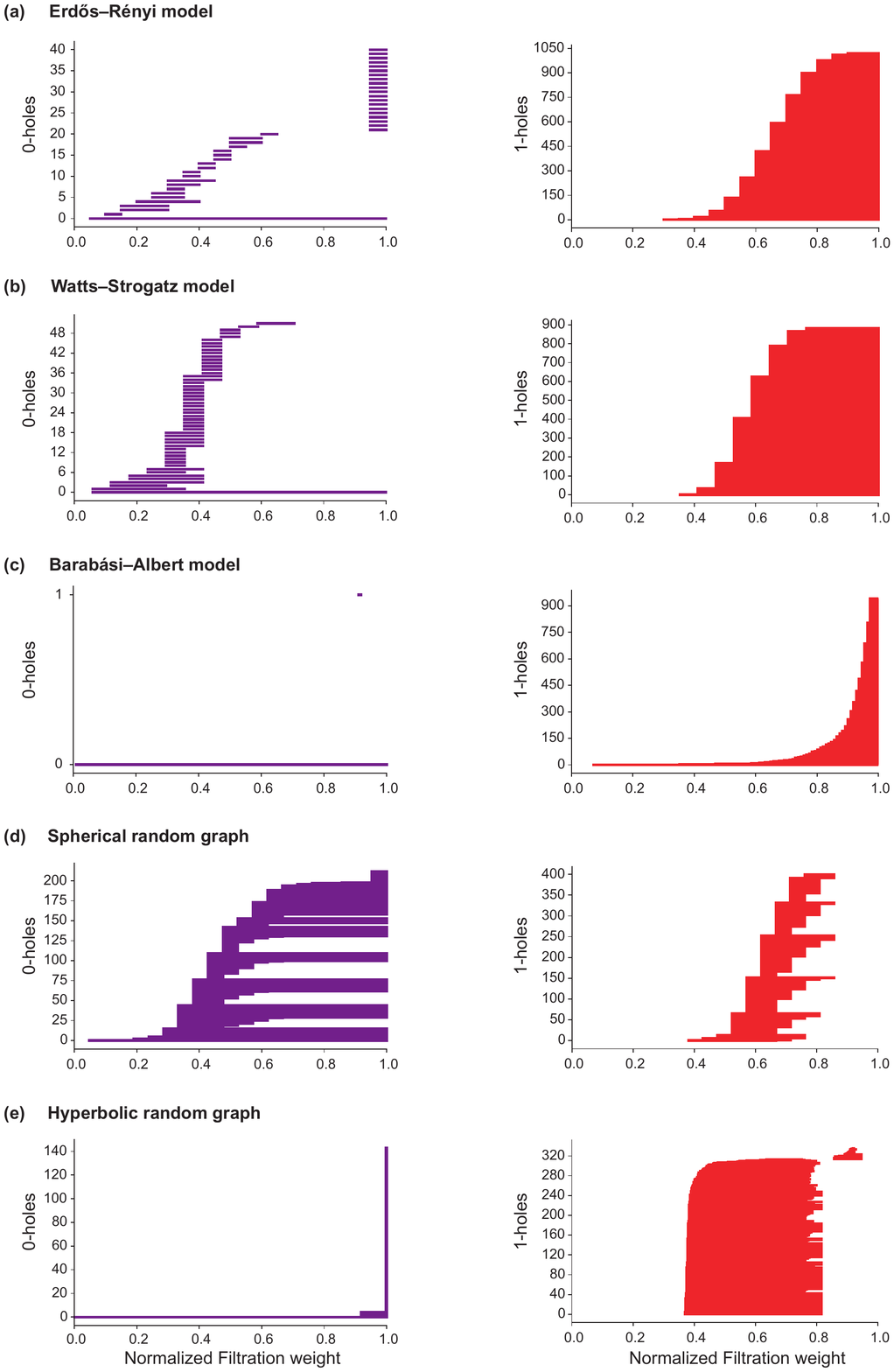}
\caption{$H_0$ and $H_1$ barcode diagrams obtained using our new method based on Forman-Ricci
curvature in model networks with average degree 4. (a) ER model with $n=1000$ and $p=0.004$.
(b) WS model with $n=1000$, $k=4$ and $p=0.5$. (c) BA model with $n=1000$ and $m=2$.
(d) Spherical random graphs produced from HGG model with $n=1000$, $T=0$, $k=4$ and
$\gamma=\infty$. (e) Hyperbolic random graphs produced from HGG model with $n=1000$, $T=0$,
$k=4$ and $\gamma=2$.}
\label{fig2}
\end{figure*}

\begin{figure*}
\includegraphics[width=.7\columnwidth]{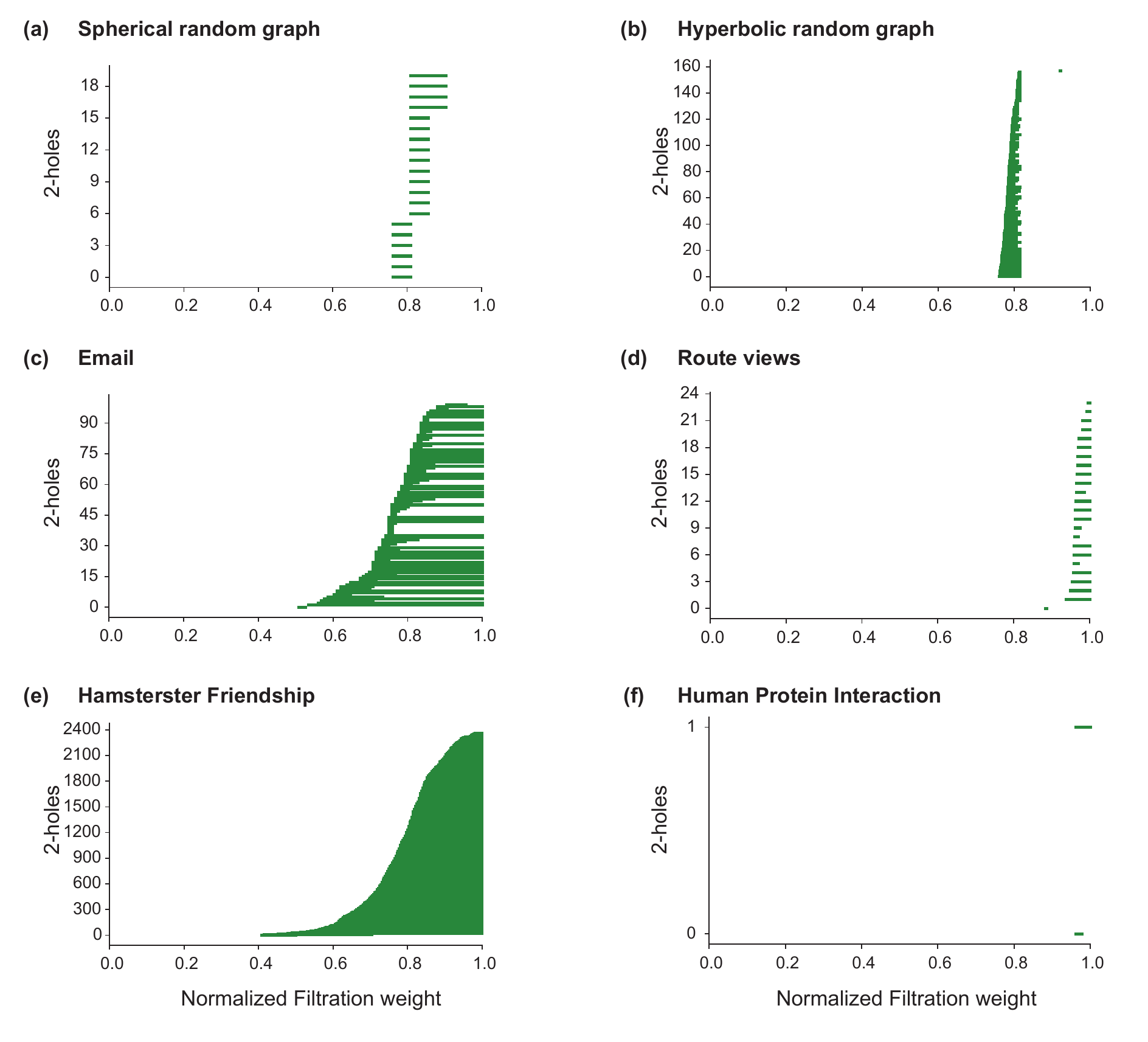}
\caption{$H_2$ barcode diagrams obtained using our new method based on Forman-Ricci curvature in
model and real networks. (a) Spherical random graph produced from HGG model with $n=1000$, $T=0$,
$k=4$ and $\gamma=\infty$. (b) Hyperbolic random graph produced from HGG model with $n=1000$,
$T=0$, $k=4$  and $\gamma=2$. (c) Email communication. (d) Route views. (e) Hamsterster friendship.
(f) Human protein interaction.}
\label{fig3}
\end{figure*}

\begin{figure*}
\includegraphics[width=.7\columnwidth]{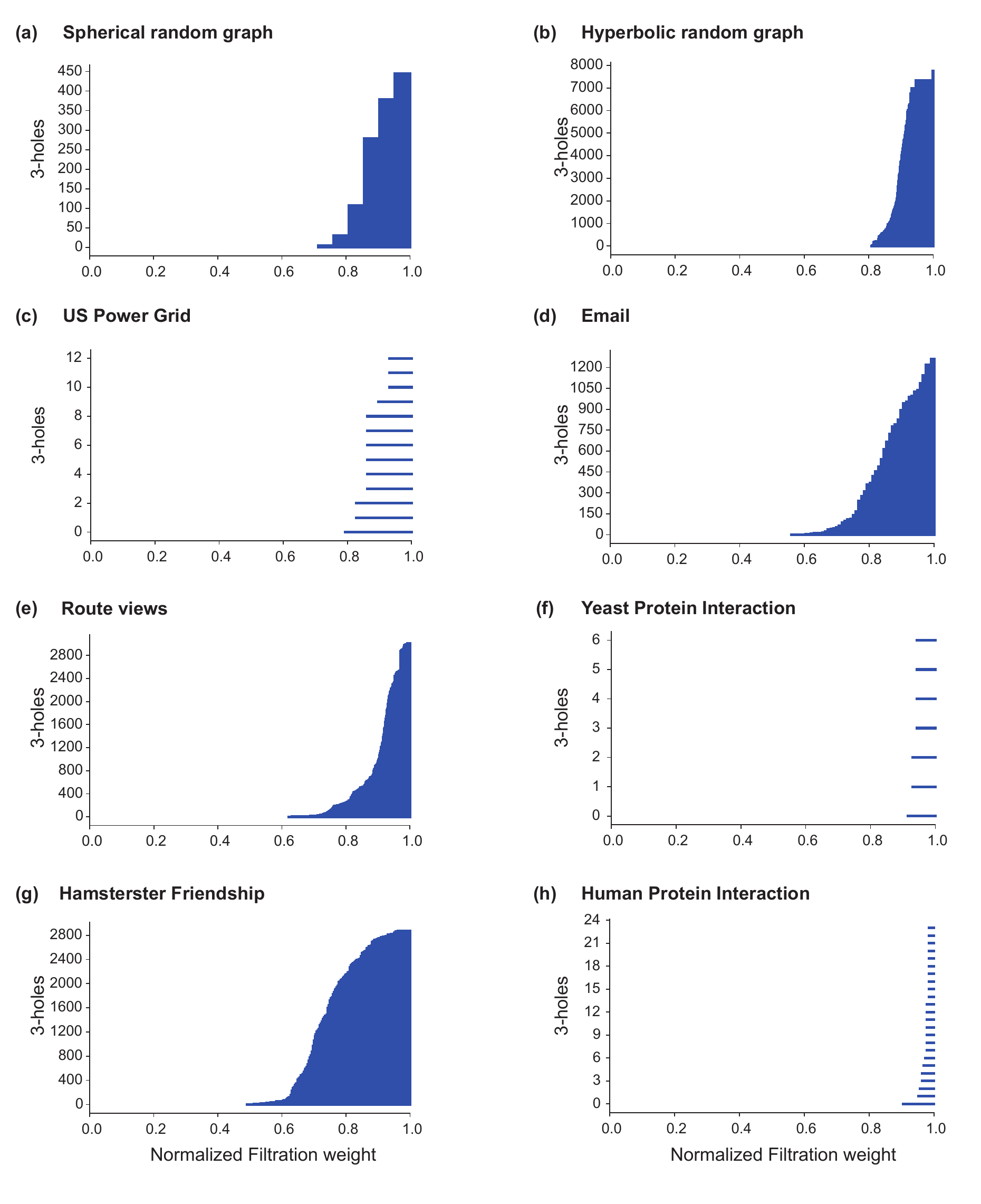}
\caption{$H_3$ barcode diagrams obtained using our new method based on Forman-Ricci curvature in
model and real networks. (a) Spherical random graph produced from HGG model with $n=1000$, $T=0$,
$k=4$ and $\gamma=\infty$. (b) Hyperbolic random graph produced from HGG model with $n=1000$,
$T=0$, $k=4$  and $\gamma=2$. (c) US Power Grid. (d) Email communication. (e) Route views.
(f) Yeast protein interaction. (g) Hamsterster friendship. (h) Human protein interaction.}
\label{fig4}
\end{figure*}

\begin{figure*}
\includegraphics[width=.5\columnwidth]{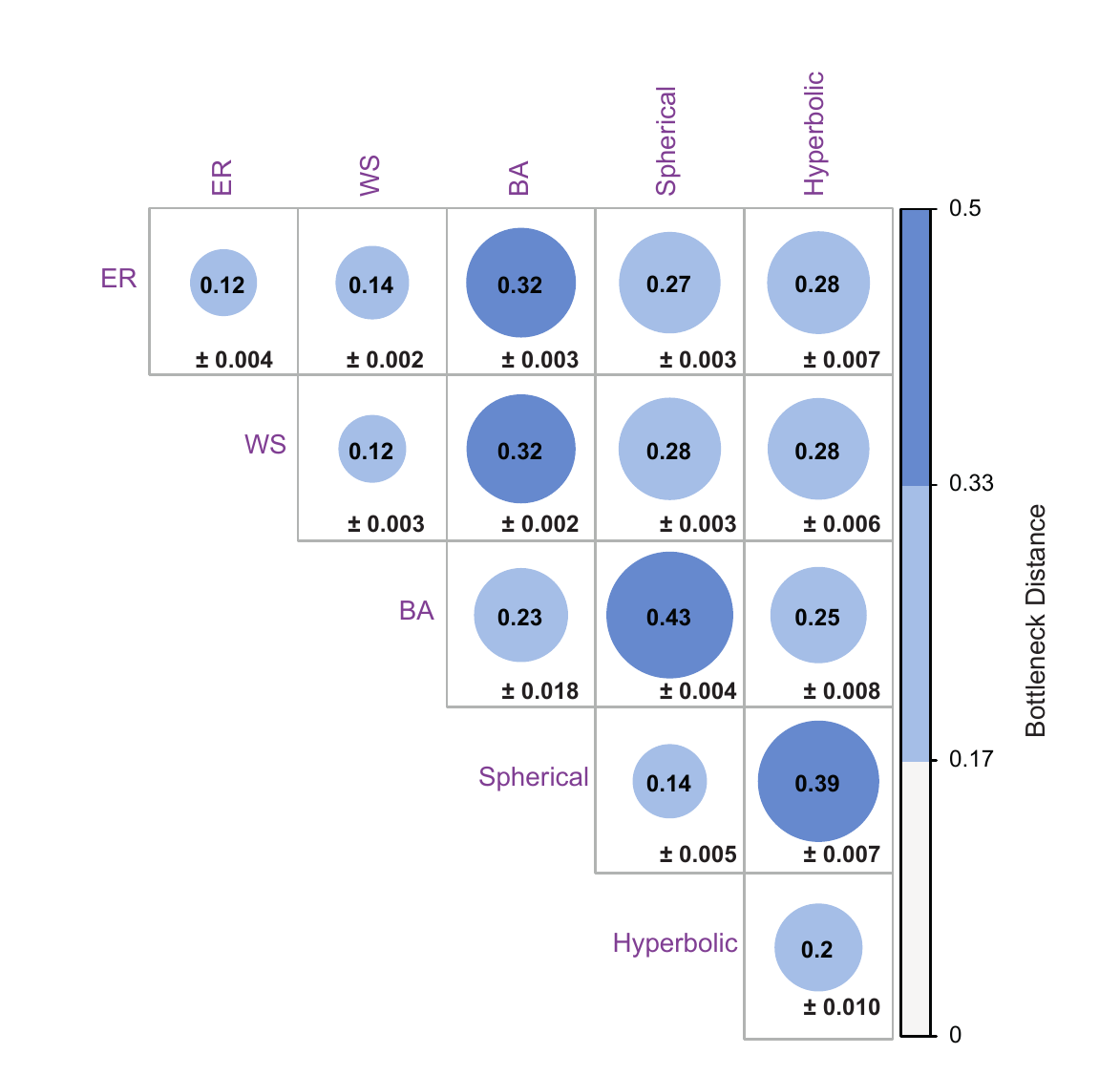}
\caption{Bottleneck distance between persistence diagrams obtained using our new method based
on Forman-Ricci curvature in model networks with average degree 4. For each of the five model
networks, 10 random samples are generated by fixing the number of vertices $n$ and other
parameters of the model. We report the distance (rounded to two decimal places) between two
different models as the average of the distance between each of the possible pairs of the 10
sample networks corresponding to the two models along with the standard error.}
\label{fig5}
\end{figure*}

\begin{figure*}
\includegraphics[width=.7\columnwidth]{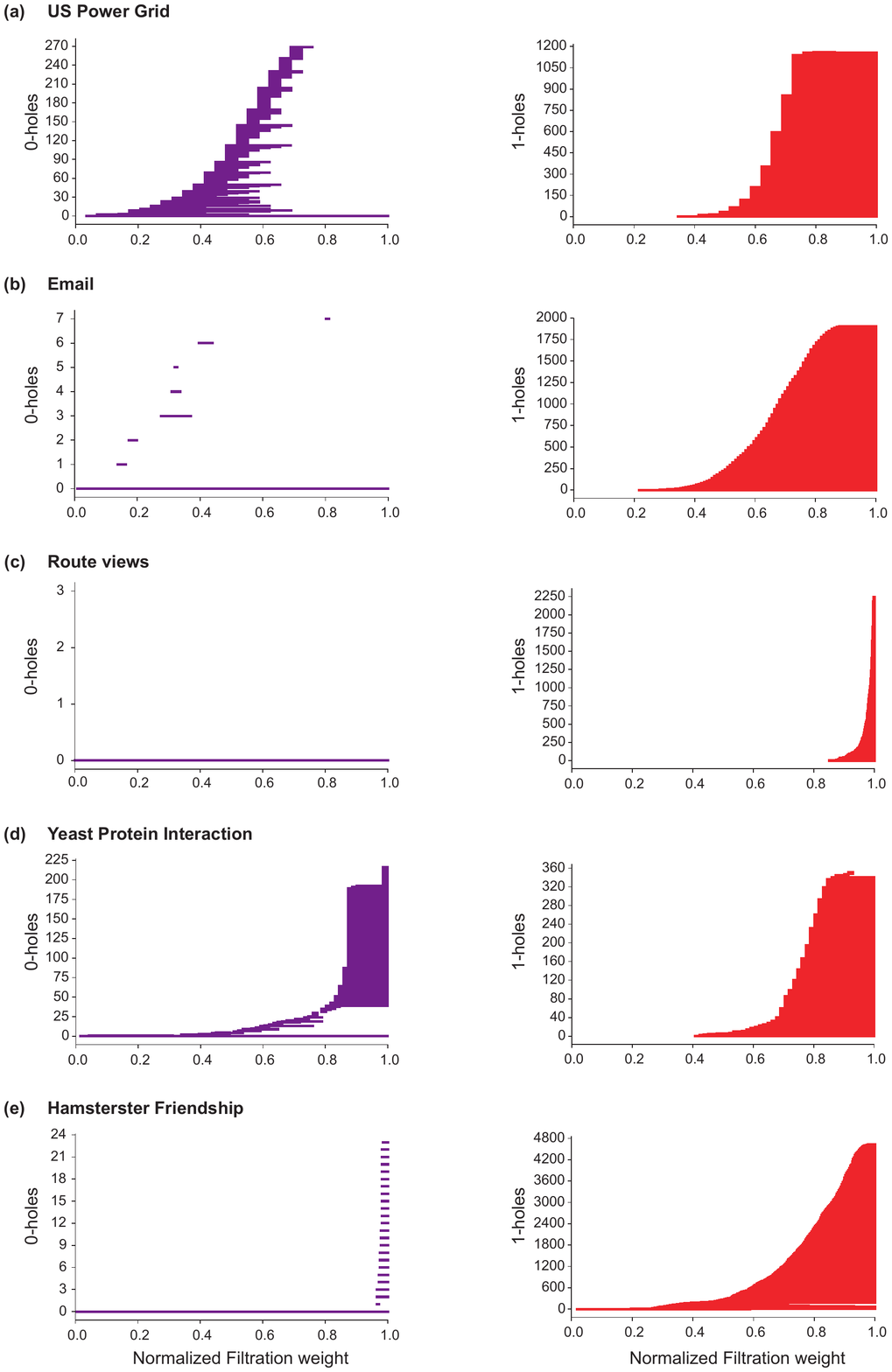}
\caption{$H_0$ and $H_1$ barcode diagrams obtained using our new method based on Forman-Ricci
curvature in real networks. (a) US Power Grid. (b) Email communication. (c) Route views. (d)
Yeast protein interaction. (e) Hamsterster friendship. }
\label{fig6}
\end{figure*}

\section{Results and Discussion}

\subsection{Persistent homology of unweighted networks using Forman-Ricci curvature}

We here present a new method based on Forman-Ricci curvature \cite{Sreejith2016,Sreejith2017,Samal2018}
to study persistent homology in unweighted and undirected networks. Essentially, our method relies on
transforming an unweighted and undirected graph $G$ into an edge-weighted network followed by
construction of a weighted clique simplicial complex $K$.

We begin by transforming a given unweighted and undirected network into an edge-weighted network by
assigning weights to edges based on their Forman-Ricci curvature (Figure \ref{schemfig}). The
Forman-Ricci curvature of each edge in an unweighted network can be computed using Equation
\ref{AugmentedFormanRicciEdge} as described in the Theory section. Thereafter, we assign weights
to edges in the network by normalization of the associated Forman-Ricci curvatures using the
following formula:
\begin{equation}
\label{FormanWeight}
w(e) = \frac { \rm{F}(e) - ( \rm{F}_{\text{min}} - \epsilon)} { ( \rm{F}_{\text{max}} + \epsilon ) -
( \rm{F}_{\text{min}} - \epsilon ) }
\end{equation}
where $w(e)$ is the weight of edge $e$, $\rm{F}(e)$ is the Forman-Ricci curvature of edge $e$,
$\rm{F}_{\text{min}}$ and $\rm{F}_{\text{max}}$ are the minimum value and maximum value, respectively,
of the Forman-Ricci curvature across all edges in the network, and $\epsilon$ is a positive number
which is taken here to be 1. In sum, the above formula gives the weights of edges in the weighted
network corresponding to the given unweighted and undirected network. In schematic Figure \ref{schemfig},
we show this transformation of an unweighted and undirected graph into an edge-weighted network using
an example.

Next, we construct a weighted clique simplicial complex starting from the edge-weighted network as
follows (Figure \ref{schemfig}). The $1$-simplices or edges in the clique complex corresponding to
the edge-weighted network already have normalized weights based on their Forman-Ricci curvature. Based on
the weights of $1$-simplices or edges, we assign weights to $0$-simplices or vertices such that the weight
of a vertex in the clique complex is equal to the minimum of the weights of edges incident on the vertex
(Equation \ref{simplexweight}). In other words, the weight of the $0$-simplex is the minimum of the
weights of its $1$-dimensional co-faces in the clique complex. Similarly, we assign weights to
$2$-simplices such that the weight of a $2$-simplex in the clique complex is equal to the maximum of the
weights of its $1$-dimensional faces or edges (Equation \ref{simplexweight}). In the same manner, we can
assign weights to higher-dimensional simplices (see Theory section). For example, the weight of a
$3$-simplex in the clique complex is equal to the maximum of the weights of its $2$-dimensional faces.
In schematic Figure \ref{schemfig}, we show this construction of a weighted clique simplicial complex
starting from an edge-weighted network using an example.

To construct a weighted clique simplicial complex corresponding to an unweighted and undirected network,
our scheme hinges on assignment of weights to $0$-simplices (vertices) based on weights of their
$1$-dimensional co-faces (i.e., edges attached to vertices). In many real-world networks, there are
isolated vertices which are not attached to any edges in the graph. In our scheme, isolated vertices
($0$-simplices) are assigned weights equal to the maximum of the weights given to any simplex in the
clique complex. In the example shown in schematic Figure \ref{schemfig}, we assign weight to the isolated
vertex $v_9$ in the weighted clique complex as described above.

After constructing the weighted clique complex $K$ corresponding to a given unweighted and undirected graph
$G$, we investigate the persistent homology of the simplicial complex via the associated filtration described
in Equation \ref{filtration} in the Theory section. In order to construct this filtration of the weighted
clique complex $K$, the assigned weights $w(\alpha_i)$ to simplices $\alpha_i$ in $K$ are arranged in an
increasing order, say $\lambda_1 \leq \lambda_2 \leq \cdots \leq \lambda_n$, and thereafter, the sequence of
subcomplexes, $K(\lambda_1) \subseteq K(\lambda_2) \subseteq \cdots \subseteq K(\lambda_n)$ is used to
compute the persistent homology groups of $K$ as described in the Theory section. In schematic Figure
\ref{schemfig}, we show this filtration of the weighted clique complex corresponding to an unweighted and
undirected network using an example.

In previous work \cite{Sreejith2016,Sreejith2017,Samal2018}, it was shown that edges critical for the
robustness of a complex network have highly negative Forman-Ricci curvature. From Equation \ref{FormanWeight},
it follows that the assigned weights to edges or $1$-simplices in the weighted clique complex $K$ constructed
by our scheme is likely to be inversely proportional to their importance from robustness perspective.
Noteworthy, critical edges for the integrity of the network are likely to be added in the initial stages of the
filtration of the weighted clique complex $K$, and thus, our method for studying the persistent homology
revolves around the central idea that the more important features of the network are included in the
initial stages of filtration.

We emphasize that the proposed method summarized in Figure \ref{schemfig} to study persistent homology in
unweighted networks basically relies on transforming an unweighted network into an edge-weighted graph which
is then used to construct a weighted clique complex. In principle, an edge-weighted graph can be obtained
from an unweighted graph by assigning weights to edges based on any edge-centric measure. In our method
summarized in Figure \ref{schemfig}, we have chosen the edge-centric measure, Forman-Ricci curvature, for
this transformation. Another possible and attractive choice of an edge-centric measure for this transformation
is the edge betweenness centrality \cite{Freeman1977,Girvan2002,Newman2010}.

In this work, we have also explored this alternate choice of edge betweenness centrality to construct
edge-weighted networks and study the persistent homology of unweighted networks. The edge betweenness
centrality of an edge in an unweighted network can be computed using Equation \ref{EBC} as described in
the Theory section. Thereafter, we can assign weights to edges in the network by normalization of the
associated edge betweenness centralities using the following formula:
\begin{equation}
\label{EBCWeight}
w(e) = \frac { ( \rm{EBC}_{\text{max}} + \epsilon) - \rm{EBC}(e)} { ( \rm{EBC}_{\text{max}} + \epsilon ) -
( \rm{EBC}_{\text{min}} - \epsilon ) }
\end{equation}
where $w(e)$ is the weight of edge $e$, $\rm{EBC}(e)$ is the edge betweenness centrality of edge $e$,
$\rm{EBC}_{\text{min}}$ and $\rm{EBC}_{\text{max}}$ are the minimum value and maximum value, respectively,
of the edge betweenness centrality across all edges in the network, and $\epsilon$ is a positive number
which is taken here to be 1. Since the edges with high edge betweenness centrality are highly critical for
information flow in the network, the above equation assigns lower weights to such critical edges to ensure
their addition during initial stages of the filtration.

In the main text of this paper, we report results from the investigation of persistent homology in
unweighted networks using our method based on Forman-Ricci curvature. In the supplementary information
(SI) of this paper, we report results from the investigation of persistent homology in unweighted networks
using our method based on alternate choice of edge betweenness centrality. In the sequel, we will show
that the qualitative and quantitative results obtained in unweighted model and real networks using our
method based on Forman-Ricci curvature is very similar to our method based on edge betweenness centrality.
However, the calculation of edge betweenness centrality requires computing all shortest paths between every
distinct pair of vertices in the network, and thus, it is much more computationally expensive than Forman-Ricci
curvature. Therefore, our method based on Forman-Ricci curvature is a better choice from a computational
perspective to study persistent homology in unweighted and undirected networks.


\subsection{Implementation in model and real networks}

In this work, we have investigated five model networks and seven real-world networks
using our methods based on Forman-Ricci curvature and edge betweenness centrality described in the
preceding section to study persistent homology in unweighted and undirected networks.

For a given unweighted and undirected network $G$, either model or real-world, we first
construct the corresponding edge-weighted network based on Forman-Ricci curvature
(Equation \ref{FormanWeight}) or edge betweenness centrality (Equation \ref{EBCWeight});
thereafter, we construct a weighted clique simplicial complex $K$ and then study the
corresponding filtration based on the edge weights as described in the preceding section.
We remark that our investigation of the persistent homology in model and real networks
is limited to the $3$-dimensional clique simplicial complex $K$ corresponding to $G$.
That is, we only include $p$-simplices which have $0 \leq p \leq 3$ while constructing the
weighted clique simplicial complex $K$ starting from an unweighted graph $G$. For these
computations of persistent homology in model and real networks, we use GUDHI \cite{Maria2014}
which is a C++ based library for Topological Data Analysis (http://gudhi.gforge.inria.fr/).

In the following, we present our results from application of our method based on Forman-Ricci
curvature to study persistent homology in model and real networks (Figure \ref{schemfig}).
We have studied here five different model networks, namely, ER, WS, BA, spherical random graphs
and hyperbolic random graphs (see Datasets section). The $0$-holes or $H_0$ barcode diagram gives
the number of connected components in the network at every stage of the filtration. We find that
the ER and WS networks possess a large number of components throughout the filtration in comparison
to BA networks where there is typically a single component which persists across the entire filtration
(Figure \ref{fig2}, SI Figures S1 and S2). This suggests that the simplices which are critical for
the overall connectivity of the network are introduced during the initial stages of the filtration
of BA networks in contrast to ER and WS networks. The $H_1$ barcode diagram indicates the presence of
$1$-holes in the network. We find that the $1$-holes appear earlier during filtration in ER and WS
networks in comparison to BA networks where $1$-holes appear towards the end of filtration (Figure
\ref{fig2}, SI Figures S1 and S2). Thus, the $H_0$ and $H_1$ barcode diagrams are able to
qualitatively distinguish scale-free BA networks from random ER networks and small-world WS networks
(Figure \ref{fig2}, SI Figures S1 and S2). Lastly, $H_2$ and $H_3$ barcode diagrams do not provide
any insight into the structure of ER, WS and BA networks due to the lack of $2$-holes and $3$-holes.

The ER, WS and BA networks can be distinguished from both spherical and hyperbolic random graphs
based on significantly larger number of $0$-holes or connected components in the later (Figure
\ref{fig2}, SI Figures S1 and S2). Though the number of components in the spherical and hyperbolic
random graphs are similar, the patterns of filtration sequence is a distinguishing factor. The
$0$-holes in the spherical random graph are more distributed across the filtration sequence, while
there are very few $0$-holes in the hyperbolic random graph for most part of the filtration with
the introduction of a large number of $0$-holes just before the end of the filtration. This can be
understood by the presence of many isolated vertices in the hyperbolic random graphs which are
assigned maximum weight in the corresponding weighted clique complex, and thus, appear at the end
of filtration (Figure \ref{fig2}, SI Figures S1 and S2). Moreover, $1$-holes and $2$-holes
appear during intermediate stages of filtration in both spherical and hyperbolic random graphs,
however, these holes do not persist till the end of filtration (Figures \ref{fig2}-\ref{fig3}, SI
Figures S1-S3). Noteworthy, there are many more $3$-holes in hyperbolic random graphs in comparison
to spherical graphs (Figure \ref{fig4}, SI Figure S3). Overall, these features enable qualitative
distinction between hyperbolic and spherical model graphs with very different underlying geometry.

The barcode diagrams for the five model networks obtained from our method based on Forman-Ricci
curvature can be used to make a qualitative distinction between different types of networks (Figure
\ref{fig2}-\ref{fig4}, SI Figures S1-S3). For a quantitative distinction between the topological
features of the five model networks, we use the bottleneck distance between total persistence
diagrams obtained from our method based on Forman-Ricci curvature as described in the Theory section. From Figure \ref{fig5}, it is seen that the bottleneck distance between
BA and ER or WS networks of similar average degree is higher than the distance between ER and WS
networks. Furthermore, the bottleneck distance is high between spherical and hyperbolic random
graphs of similar average degree. Overall, the bottleneck distances between the total persistence
diagrams for the five model networks provide a quantitative validation of the applicability of our
method based on Forman-Ricci curvature to reveal distinct topological features in unweighted and
undirected networks.

We have also studied here seven well-known real-world networks (see Datasets section). From the $H_0$
barcode of the US Power Grid network, it is clear that there is one connected component which persists
across the entire filtration despite many transient components appearing and disappearing during
intermediate stages of filtration (Figure \ref{fig6}). The $H_0$ barcode diagrams of the E-mail and
Route views networks are similar in the sense that the number of connected components across the
entire filtration is low (Figure \ref{fig6}). The $H_0$ barcode diagrams of the two biological networks,
namely Human protein interaction and Yeast protein interaction show a sudden increase in the number of
connected components at the final stage of filtration (Figure \ref{fig6}, SI Figure S4). The $H_0$
barcode diagram of the Euro road network displays a distributed pattern with components spanning across
varied intervals of filtration (SI Figure S4). The $H_1$ barcode diagrams of the seven real networks are
similar in the respect that there is typically an increase in the number of $1$-holes around the middle
to later stages of filtration (Figure \ref{fig6}, SI Figure S4). We also display the $H_2$ and $H_3$
barcode diagrams for real networks in Figures \ref{fig3} and \ref{fig4}, respectively. Note that the
Euro road network is devoid of $2$-holes and $3$-holes, while the Yeast protein interaction network
and US Power Grid network are devoid or $2$-holes (Figures \ref{fig3} and \ref{fig4}). In sum, the
barcode diagrams obtained using our method based on Forman-Ricci curvature can be used to reveal
differences between model and real networks.

In the above paragraphs, we reported our results from application of our method based on Forman-Ricci
curvature to study persistent homology in unweighted networks, both model and real-world. As described
in the preceding section, we can also apply an alternate method based on edge betweenness centrality
to study persistent homology in unweighted networks, both model and real-world. In SI Figures S5-S9,
we present the $H_0$, $H_1$, $H_2$ and $H_3$ barcode diagrams obtained using our alternate method based
on edge betweenness centrality in five model and seven real-world networks. Based on SI Figures S5-S9,
it is evident that we are also able to make qualitative distinction between varied model networks using
the barcode diagrams obtained from our method based on edge betweenness centrality, and these results are
similar to results described above from our method based on Forman-Ricci curvature. Moreover, in SI
Figure S10, we display the bottleneck distances between persistence diagrams corresponding to the five
model networks obtained using the alternate method based on edge betweenness centrality, and these
results are also similar to those shown in Figure \ref{fig5} which are obtained using our method based
on Forman-Ricci curvature.


\subsection{Comparison with method based on discrete Morse theory}

Classical Morse theory on smooth manifolds has been a rich theory to detect the topology of the
underlying space \cite{Morse1934}. However, it requires a smooth structure to probe the topology
via real-valued smooth functions. Robin Forman introduced \textit{discrete Morse theory}, the
discrete counterpart of classical Morse theory \cite{Forman1998,Forman2002}, which is applicable
to a large class of topological objects called $CW$-complexes, even those which lack smoothness.
Similar to classical Morse theory, this discretized version also captures the topology of the
underlying object. A fundamental notion in discrete Morse theory is that of \textit{critical cells},
which are the discrete analogues of equilibrium points of a Morse function, i.e. points on which
its gradient vanishes. The number of such critical cells is intricately related to the Betti
numbers and the Euler characteristic of the topological space via the Morse inequalities
\cite{Forman1998,Forman2002}.

In our previous work \cite{Kannan2019}, we have used discrete Morse theory of Robin Forman to set
weights on the clique complex of an unweighted graph. We also gave an algorithm to produce a
near-optimal discrete Morse function, in the sense that the number of so-called critical simplices
of the function is close to the theoretical minimum given by Betti numbers. The advantages of this
approach lies in the ability for \textit{preprocessing} of the simplicial complex, which leads to
significant simplification that in turn, leads to computational efficiency for homology calculations
\cite{Mischaikow2013,Harker2014}. In principle, one can use this method independent of TDA, for
example to study combinatorial topology aspects of such complexes \cite{Shareshian2001}. As an
application to TDA, we \cite{Kannan2019} used this method to compute the persistent homology of the
model and real-world networks which are also analyzed in this contribution, and showed that it was
able to distinguish between various networks with different topological features.

Our present method focusses on the computation of persistent homology only, setting aside the
computational advantages of using a near-optimal discrete Morse function. We have shown that even
though we lose the simplified topological structure provided by discrete Morse theory, we can still
apply other weighting schemes using local and global measures like Forman-Ricci curvature and edge
betweenness centrality, to compute persistent homology of various networks. Indeed, the new method
also distinguishes the various model networks as well as the previous method using discrete Morse
theory. Therefore, using these new weighting schemes can be seen as a tradeoff between computational
efficiency and applicability of TDA to unweighted graphs.


\section{Conclusions}

This work is meant to provide techniques to apply TDA, for the investigation of topological properties
of unweighted networks. We employ two methods to convert an unweighted network into a weighted network.
The first one is a discretized version of Ricci curvature which takes into account only local information
around an edge in the network, while the second one is a global quantifier which measures the importance
of an edge based on the number of shortest paths between any two distinct vertices of the network passing
through that edge. Once we have a weighted graph, standard techniques allow us to obtain a filtration of
the associated clique complexes, and thereby compute their persistent homology (Figure \ref{schemfig}).
We have applied this method to study five different kinds of model networks. We also show that this method
can distinguish different model networks via the averaged bottleneck distance between the corresponding
persistence diagrams (Figure \ref{fig5}). We also apply the same techniques to study some well-known
real-world networks to obtain insights into their underlying topology. In future work, we plan to apply
these techniques to networks arising out of physical systems, with the goal of trying to find correlations
between dynamics of such systems and their underlying topology via the application of TDA.



\subsection*{Funding}

This work was supported by Max Planck Society, Germany, through the award of a Max
Planck Partner Group in Mathematical Biology (to A.S.) and Science and Engineering
Research Board (SERB), Department of Science and Technology (DST) India through the
award of a MATRICS grant [MTR/2017/000835] (to I.R.).

\subsection*{Declaration of Competing Interest}

None.

\subsection*{Acknowledgments}

S.V. and S.J.R. thank the Institute of Mathematical Sciences (IMSc), Chennai, India
for their local hospitality and R. Nadarajan for encouragement. 

\subsection*{Author contributions}

I.R. and A.S. designed the study and developed the method. S.V. and S.J.R. performed
the simulations. I.R. and A.S. analyzed results. I.R., S.V., S.J.R. and A.S. wrote
the manuscript. All authors reviewed and approved the manuscript.

%

\newpage
\begin{center}
\section*{\large \bf SUPPLEMENTARY INFORMATION (SI)}
\end{center}
\renewcommand{\theequation}{S.\arabic{equation}}
\renewcommand{\thefigure}{S\arabic{figure}}
\setcounter{equation}{0}
\setcounter{figure}{0}

\begin{figure}[hbt!]
\includegraphics[width=.7\columnwidth]{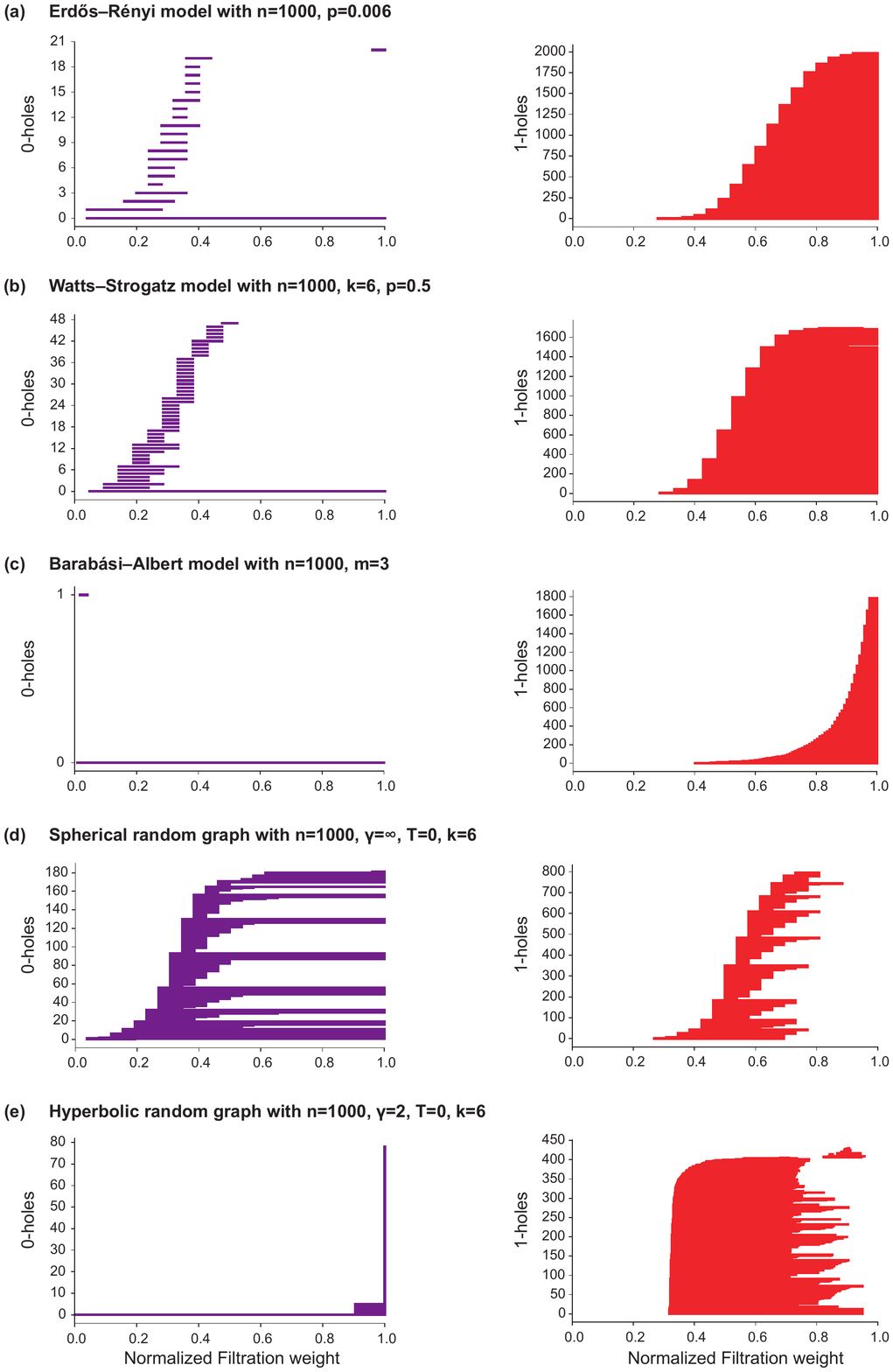}
\caption{$H_0$ and $H_1$ barcode diagrams obtained using our new method based on Forman-Ricci
curvature in model networks with average degree 6.}
\end{figure}

\begin{figure}
\includegraphics[width=.7\columnwidth]{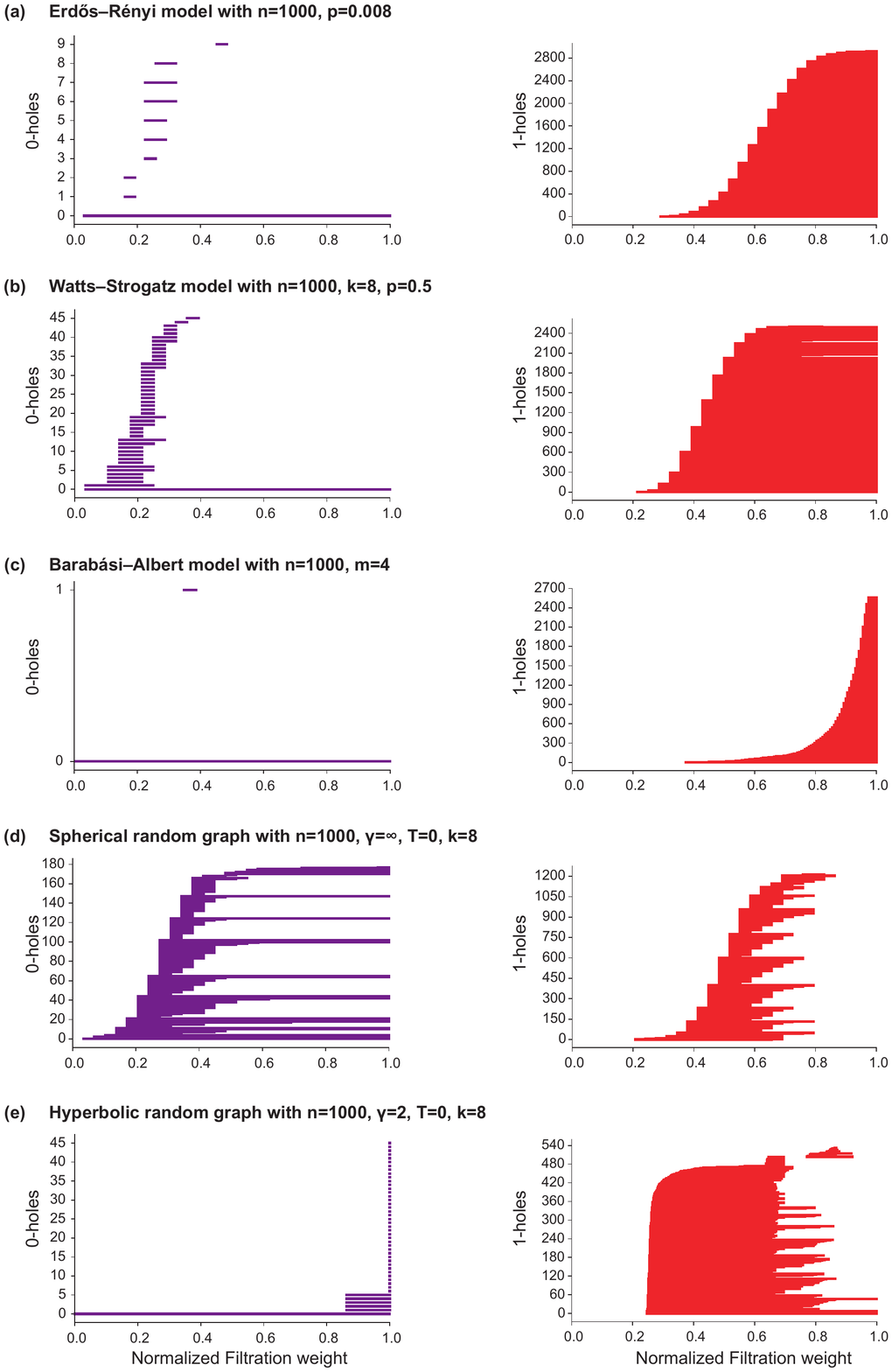}
\caption{$H_0$ and $H_1$ barcode diagrams obtained using our new method based on Forman-Ricci
curvature in model networks with average degree 8.}
\end{figure}

\begin{figure}
\includegraphics[width=.7\columnwidth]{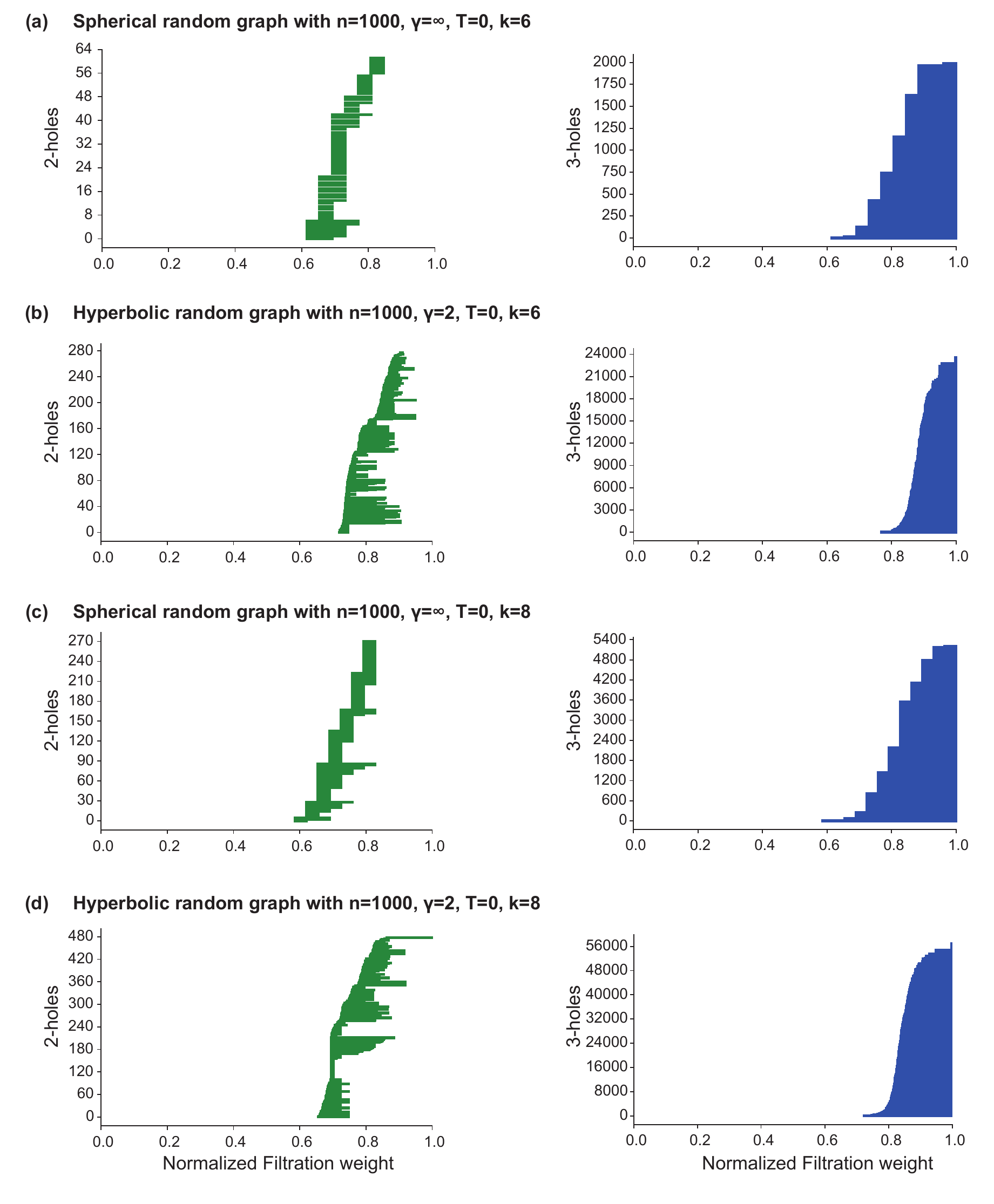}
\caption{$H_2$ and $H_3$ barcode diagrams obtained using our new method based on Forman-Ricci
curvature in model networks with average degree 6 and 8.}
\end{figure}

\begin{figure}
\includegraphics[width=.7\columnwidth]{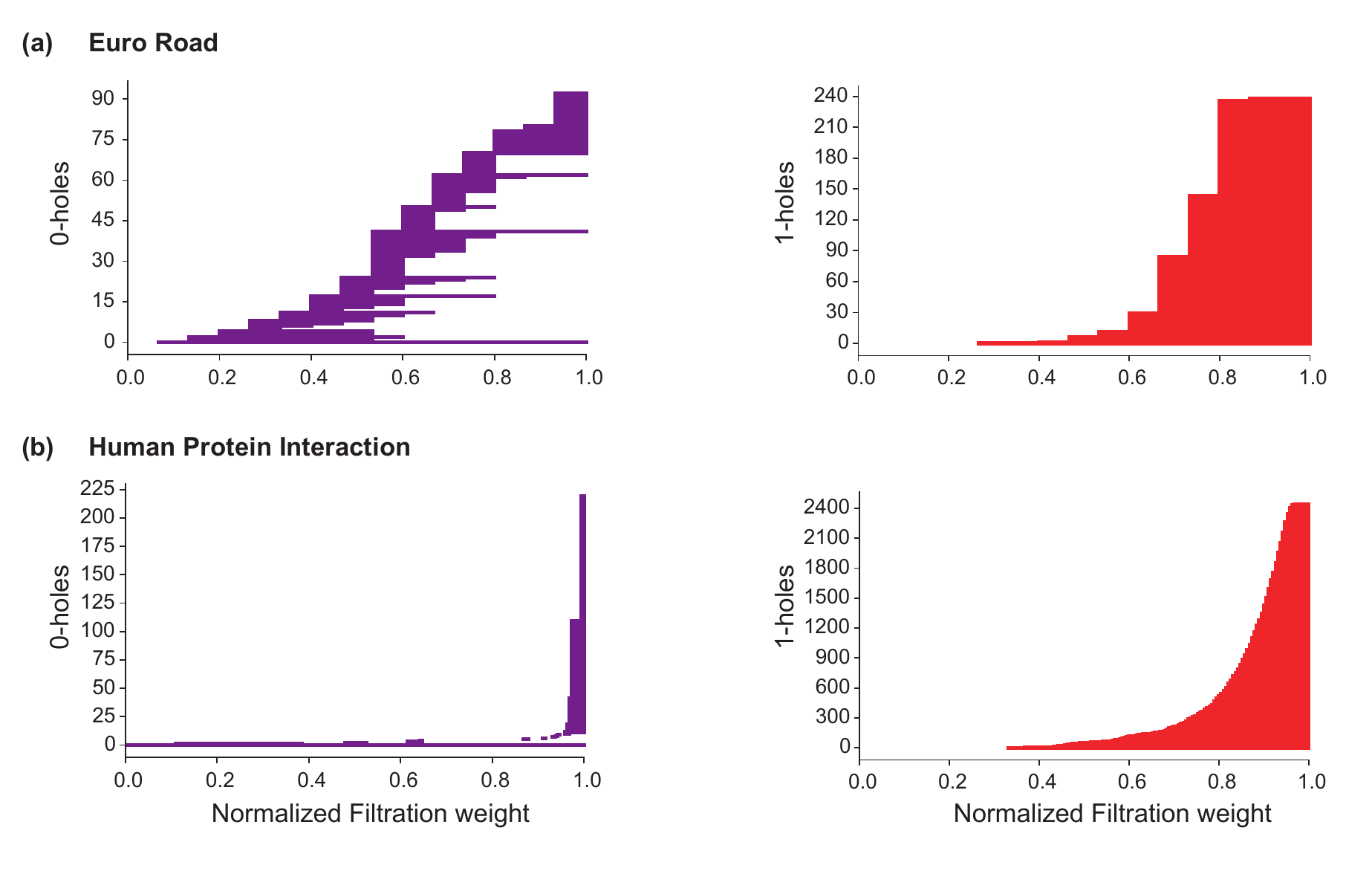}
\caption{$H_0$ and $H_1$ barcode diagrams obtained using our new method based on Forman-Ricci
curvature in real networks. (a) Euro road. (b) Human protein interaction. }
\end{figure}

\begin{figure}
\includegraphics[width=.7\columnwidth]{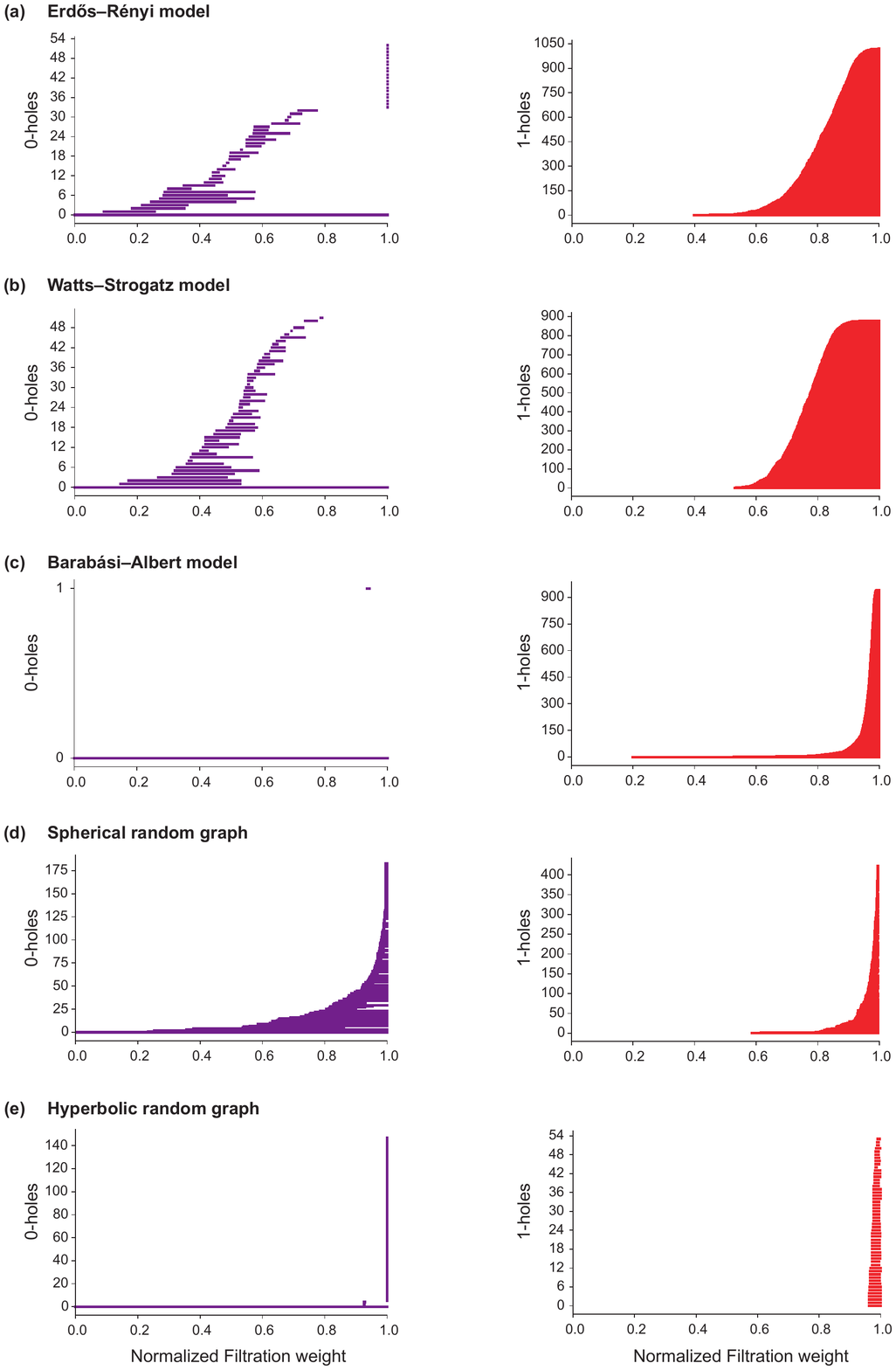}
\caption{$H_0$ and $H_1$ barcode diagrams obtained using our new method based on edge betweenness
centrality in model networks with average degree 4.}
\end{figure}

\begin{figure}
\includegraphics[width=.7\columnwidth]{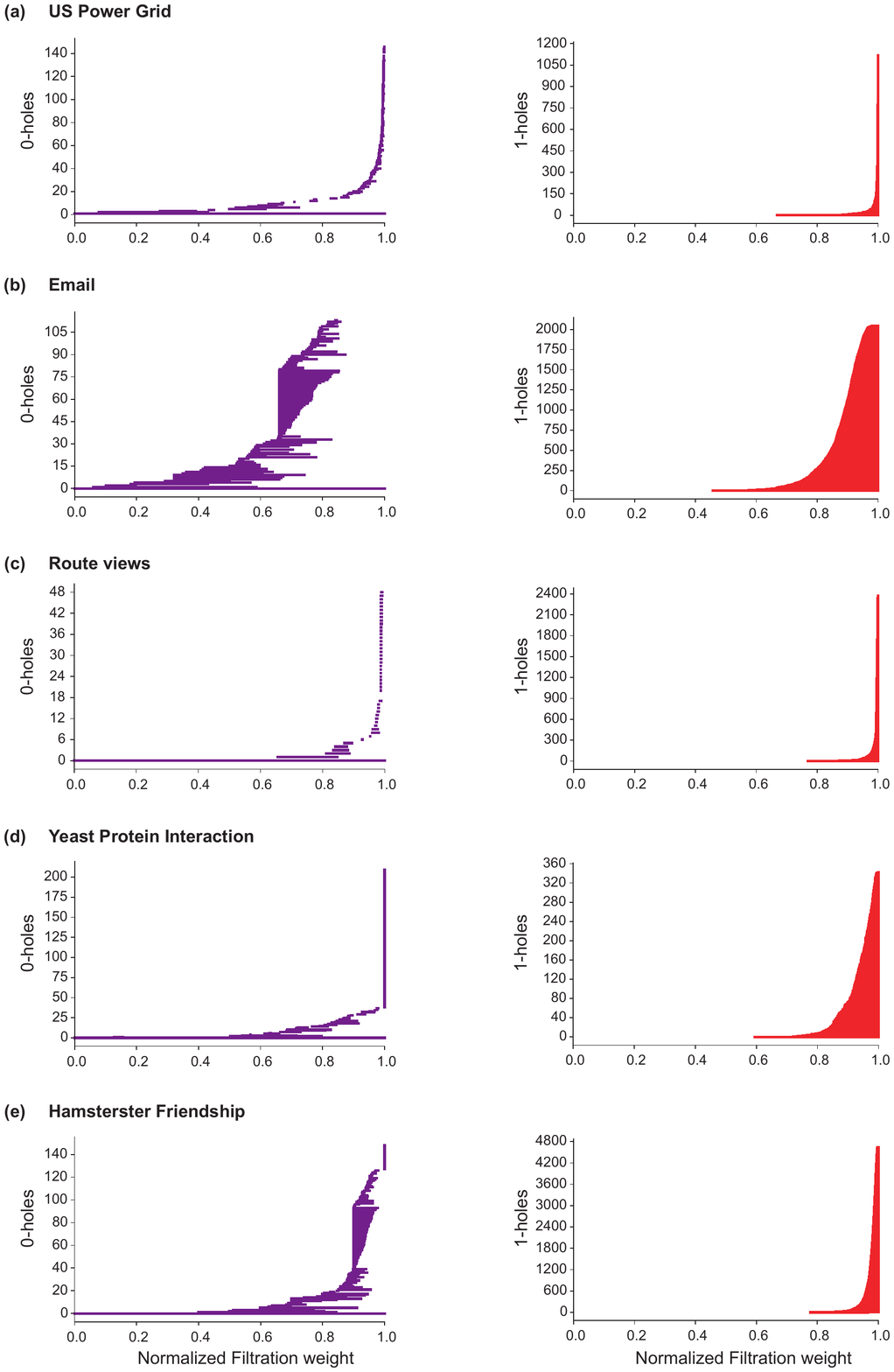}
\caption{$H_0$ and $H_1$ barcode diagrams obtained using our new method based on edge betweenness
centrality in real networks. (a) US Power Grid. (b) Email communication. (c) Route views. (d)
Yeast protein interaction. (e) Hamsterster friendship.}
\end{figure}

\begin{figure}
\includegraphics[width=.7\columnwidth]{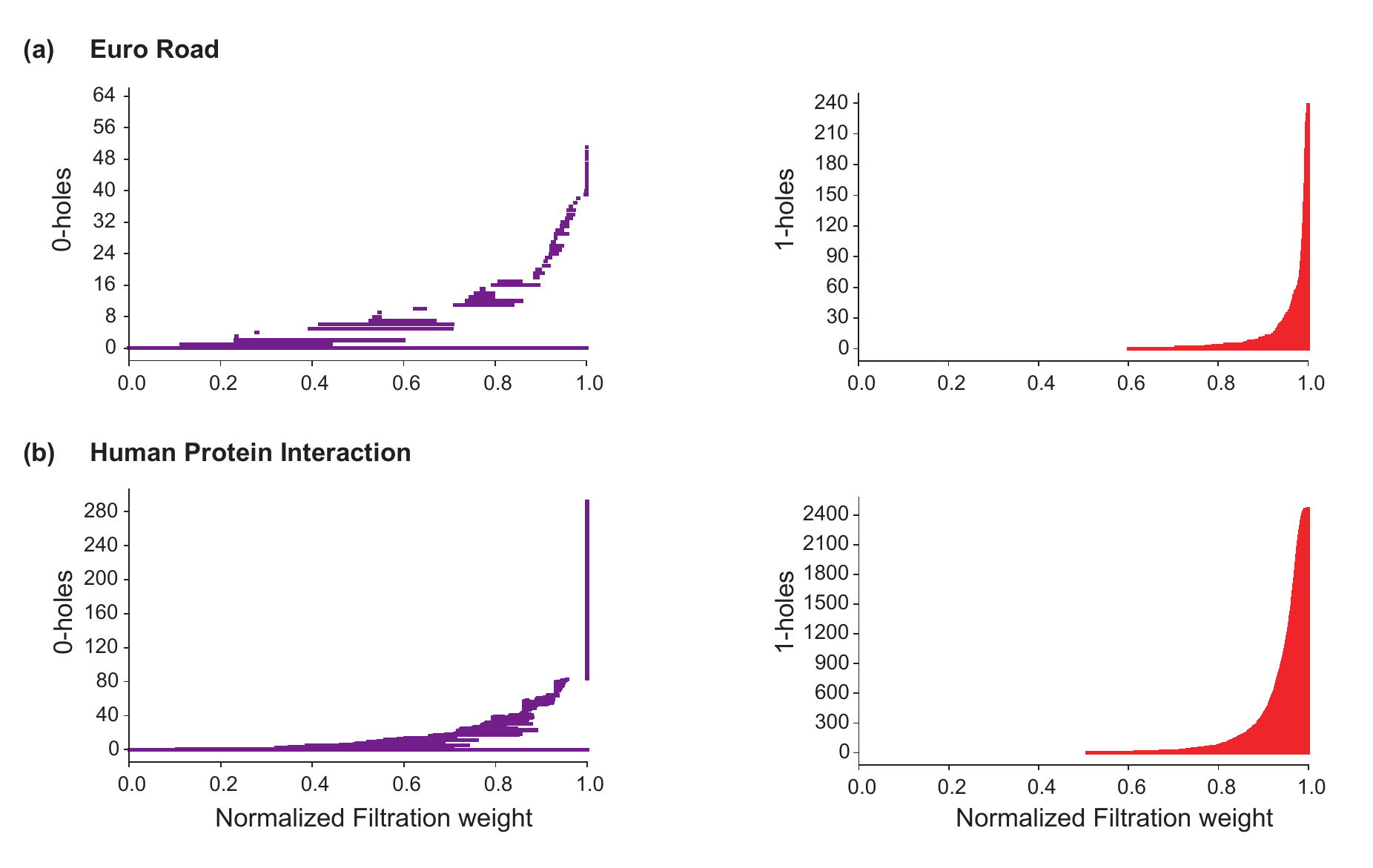}
\caption{$H_0$ and $H_1$ barcode diagrams obtained using our new method based on edge betweenness
centrality in real networks. (a) Euro road. (b) Human protein interaction. }
\end{figure}

\begin{figure*}
\includegraphics[width=.7\columnwidth]{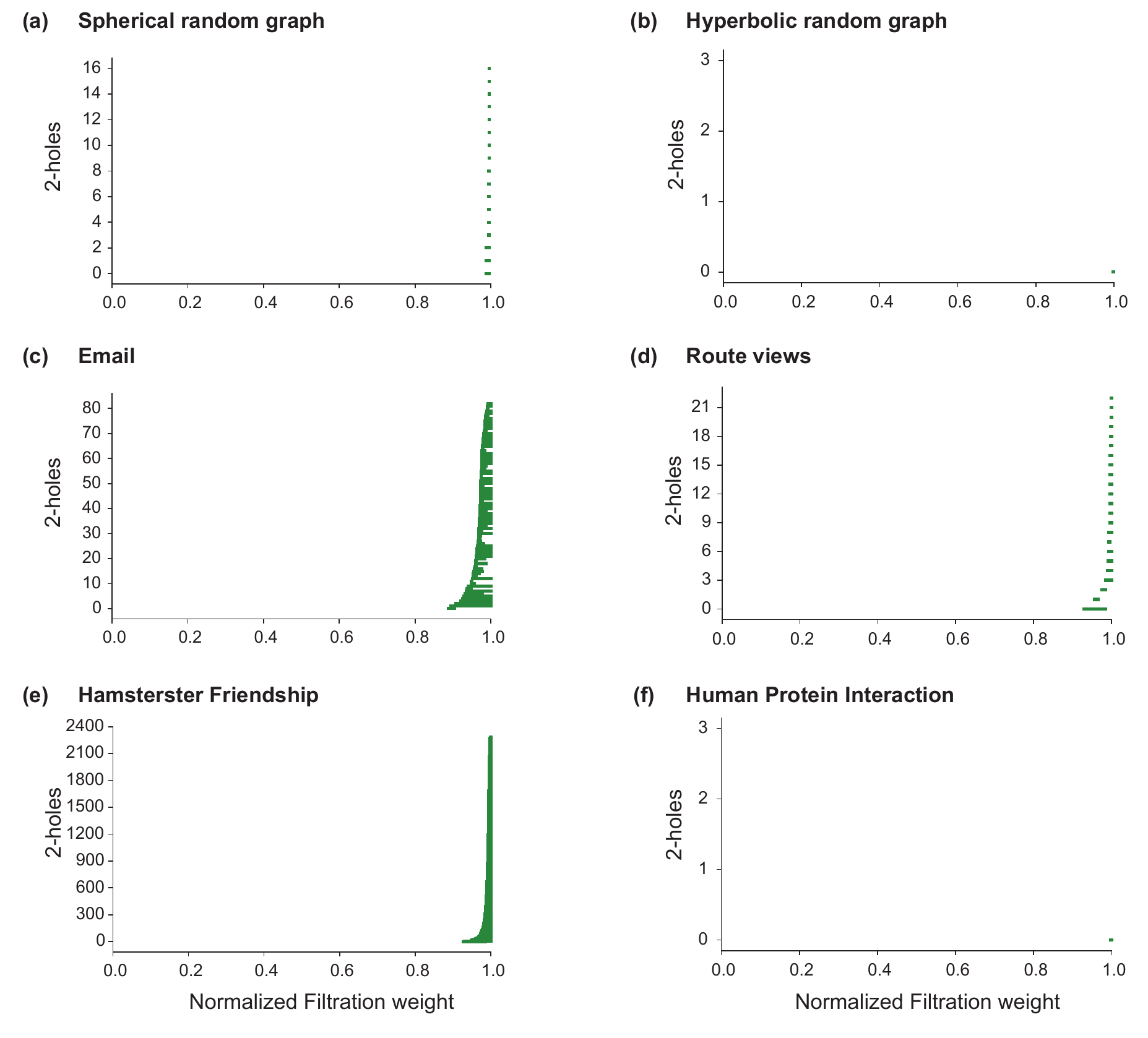}
\caption{$H_2$ barcode diagrams obtained using our new method based on edge betweenness centrality
in model and real networks. (a) Spherical random graph produced from HGG model with $n=1000$, $T=0$,
$k=4$ and $\gamma=\infty$. (b) Hyperbolic random graph produced from HGG model with $n=1000$,
$T=0$, $k=4$  and $\gamma=2$. (c) Email communication. (d) Route views. (e) Hamsterster friendship.
(f) Human protein interaction.}
\end{figure*}

\begin{figure*}
\includegraphics[width=.7\columnwidth]{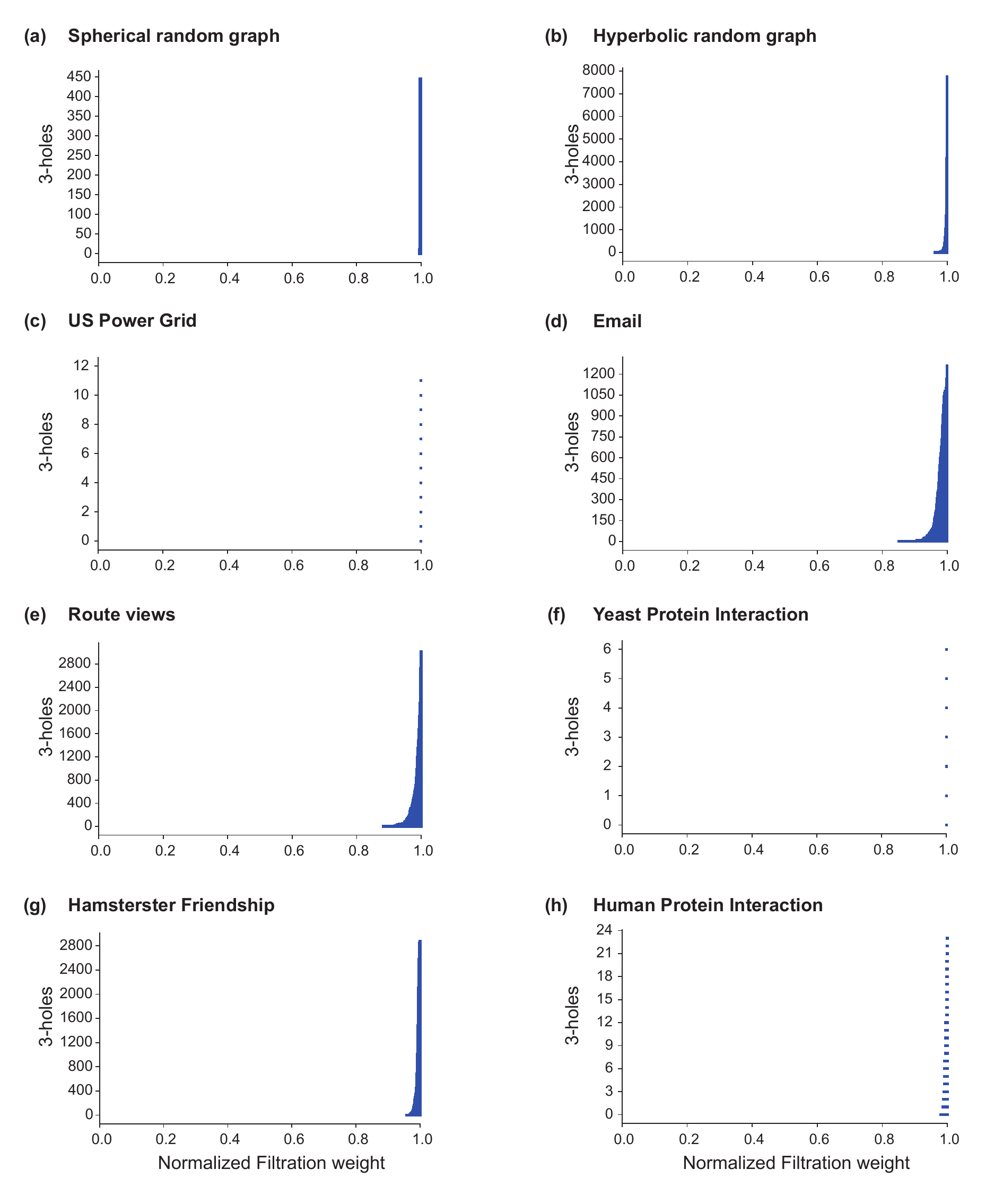}
\caption{$H_3$ barcode diagrams obtained using our new method based on edge betweenness centrality
in model and real networks. (a) Spherical random graph produced from HGG model with $n=1000$, $T=0$,
$k=4$ and $\gamma=\infty$. (b) Hyperbolic random graph produced from HGG model with $n=1000$,
$T=0$, $k=4$  and $\gamma=2$. (c) US Power Grid. (d) Email communication. (e) Route views.
(f) Yeast protein interaction. (g) Hamsterster friendship. (h) Human protein interaction.}
\end{figure*}

\begin{figure*}
\includegraphics[width=.5\columnwidth]{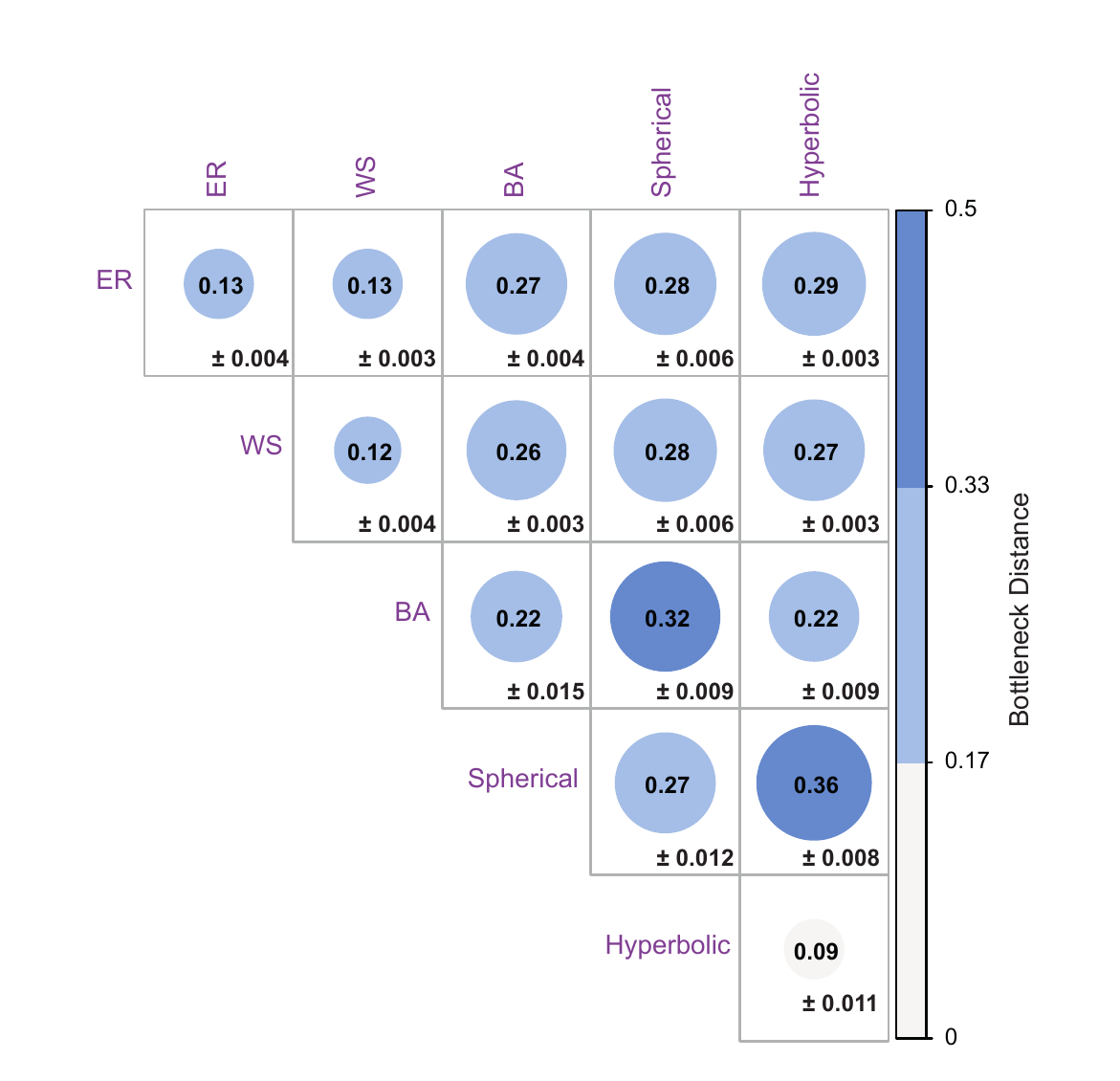}
\caption{Bottleneck distance between persistence diagrams obtained using our new method based
on edge betweenness centrality in model networks with average degree 4. For each of the five
model networks, 10 random samples are generated by fixing the number of vertices $n$ and other
parameters of the model. We report the distance (rounded to two decimal places) between two
different models as the average of the distance between each of the possible pairs of the 10
sample networks corresponding to the two models along with the standard error.}
\end{figure*}

\end{document}